\documentclass{article}

\usepackage[numbers,sort&compress]{natbib}
\usepackage{algorithm}
\usepackage{algorithmic}
\usepackage{amssymb}

\usepackage{wrapfig}

     \usepackage[preprint]{neurips_2025}

\usepackage[utf8]{inputenc} 
\usepackage[T1]{fontenc}    
\usepackage{hyperref}       
\usepackage{url}            
\usepackage{booktabs}     
\usepackage{amsfonts}       
\usepackage{nicefrac}       
\usepackage{microtype}      
\usepackage{xcolor}         

\usepackage{graphicx}
\usepackage{subfigure}
\usepackage{amsmath}
\title{Discrete Spatial Diffusion: Intensity-Preserving Diffusion Modeling}

\author{
  {\large Javier E. Santos\thanks{Los Alamos National Laboratory}} \\
  \And
  {\large Agnese Marcato\footnotemark[1]} \\
  \And
  {\large Roman Colman\footnotemark[1]} \\
  \And
  {\large Nicholas Lubbers\footnotemark[1]} \\
  \And
  {\large Yen Ting Lin\footnotemark[1]}
}

\begin{document}

\maketitle

\begin{abstract}
Generative diffusion models have achieved remarkable success in producing high-quality images. However, these models typically operate in continuous intensity spaces, diffusing independently across pixels and color channels. As a result, they are fundamentally ill-suited for applications involving inherently discrete quantities—such as particle counts or material units—that are constrained by strict conservation laws like mass conservation, limiting their applicability in scientific workflows. To address this limitation, we propose Discrete Spatial Diffusion (DSD), a framework based on a continuous-time, discrete-state jump stochastic process that operates directly in discrete spatial domains while strictly preserving particle counts in both forward and reverse diffusion processes. By using spatial diffusion to achieve particle conservation, we introduce stochasticity naturally through a discrete formulation. We demonstrate the expressive flexibility of DSD by performing image synthesis, class conditioning, and image inpainting across standard image benchmarks, while exactly conditioning total image intensity.
We validate DSD on two challenging scientific applications: porous rock microstructures and lithium-ion battery electrodes, demonstrating its ability to generate structurally realistic samples under strict mass conservation constraints, with quantitative evaluation using state-of-the-art metrics for transport and electrochemical performance.
\end{abstract}

\section{Introduction}

Diffusion-based generative models have emerged as powerful tools for high-quality image generation \cite{sohl-dicksteinDeepUnsupervisedLearning2015,hoDenoisingDiffusionProbabilistic2020,songScoreBasedGenerativeModeling2021a}. Typically, these models inject noise into the images, then learn to reverse this noise-adding process to recover meaningful structure. In most frameworks, this is based on an It\^o Stochastic Differential Equation (SDE) with Gaussian noise. While effective for many vision tasks, these approaches inherently assume continuous pixel intensities, which can cause difficulty when dealing with the discrete nature of many datasets. Tasks beyond vision, such as those in the physical sciences have many applications which require discrete physical quantities, such as particle counts in a simulation, or phases in materials microstructure. Conservation of total quantities is critical for scientific applications, and so generative modeling which can operate under constrained, discrete pixel intensities would enable scientifically  grounded and physically consistent synthesis. Such a capability might also prove useful within vision tasks, such as inpainting and super-resolution.

Scientific and engineering studies of the natural world using computational techniques often involve discrete variables in space and/or time. On microscopic scales, everyday materials exhibit extremely complex structural patterns which encode the history of their formation, and play a large role in how the material functions on a macroscopic level. An important and wide-reaching field of study is materials microstructure, which is used in materials design \cite{micromaterials}, forensic analysis, hydrology \cite{blunt2013}, energy storage \cite{Simonenergystorage}, and even medicine, such as in studies  of bone structure \cite{Montoyabone2021}.  For example, crystal grain shapes can give rise to complex stress patterns which affect the yield strength of a metal \cite{calcagnotto2011deformation}. Materials microstructures are frequently represented by a small number of discrete phases that describe their underlying chemical structures. In sandstone, for example, the overall arrangement of nanocrystals is highly disordered, and gives rise to complex pore structures, through which subsurface water flows. This microstructure has an enormous influence on the rate of transport of fluids and contaminants \cite{JacobBear}. Microstructure of electrodes is also known to have an immense impact on the characteristics of electrochemical devices \cite{phogat2024microstructural}.
Small changes in thermodynamic properties can cause drastic changes in microstructure, such as in stainless steels \cite{xiong2010phase}, requiring the study of microstructure as a function of phase contents. Furthermore, gathering real-world data is often complex and expensive; decades of work have been applied to computational modeling of the generation and consequences of microstructure prior to the widespread popularization of machine learning \cite{torquato2002random}. While diffusion models have been constructed for microstructure~\cite{Lyu2024Microstructure,hoffman2025grain,lee2024data,lee2024microstructure,bluntdiffusion,azqadan2023predictive,dureth2023conditional}, they have not handled exact porosity (intensity) constraints.

In this work, we introduce Discrete Spatial Diffusion (DSD), a discrete-state Markov chain-based diffusion framework in which the forward process redistributes discrete units of intensity in space (Fig.~\ref{fig:schematic}c). Unlike previous diffusion models, DSD  \textit{exactly} preserves total intensity throughout both the forward and reverse phases, ensuring that global properties---such as mass fractions---are exactly conserved by representing them in terms of conserved particles. We demonstrate that DSD enables scientific applications and can extend the capabilities of conventional image processing tasks, like image generation and inpainting in discrete domains. By directly modeling discrete transitions, DSD enables generative modeling under conservation laws, allowing models that specialize for constrained conditions in scientific applications and beyond. 

\section{Background}

Among the body of literature on generative diffusion models, originating from the pioneering work of \citet{sohl-dicksteinDeepUnsupervisedLearning2015}, the most relevant to our work fall into two broad categories: (1) those employing discrete-state Markov chains to introduce noise in the forward process \citep{hoogeboom2021argmax,austin21,campbell2022,blackout,sunScorebasedContinuoustimeDiscrete2022,louDiscreteDiffusionLanguage2023,shi2024simplified}, and (2) those incorporating spatial dynamics into the forward diffusion process \citep{Bansal2022ColdNoise,heat_eq,hoogeboom2021argmax}.

\begin{figure*}[ht!]
    \centering
    \includegraphics[width=1\linewidth]{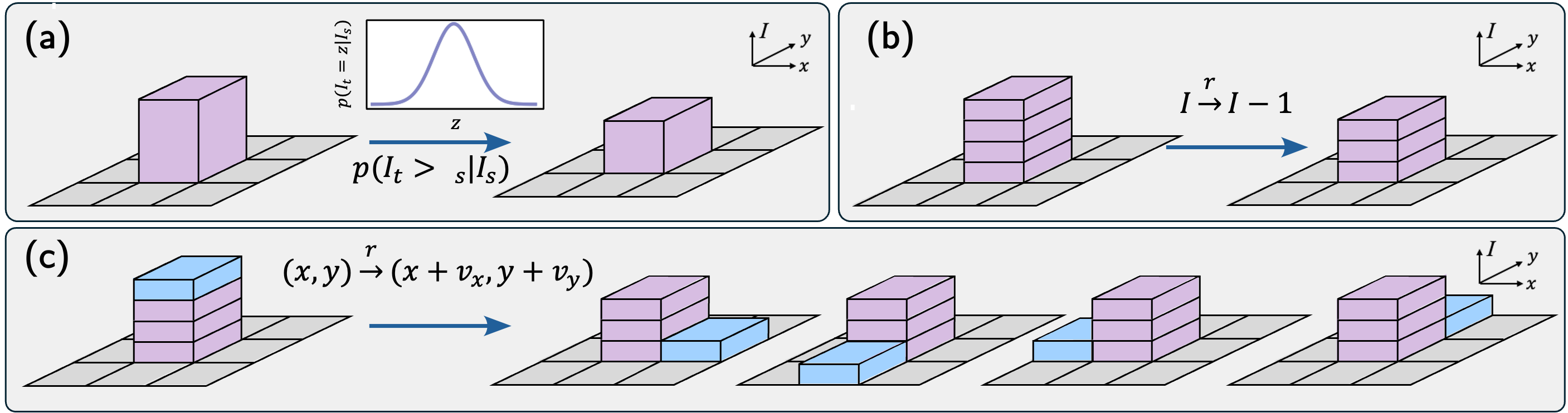}
    \vspace{-6.5mm}
    \caption{Schematic diagrams illustrating how intensity is modeled in different diffusion frameworks.  \textbf{(a)} Gaussian Diffusion
    relies on the Ornstein--Uhlenbeck  process in continuous intensity space. \textbf{(b)} Prior discrete-state diffusion models
    apply a discrete-state Markov process to independent pixel intensities. 
     \textbf{(c)} Discrete Spatial Diffusion (this work) relies on a Markov jump process of the intensity units over a discrete spatial lattice, redistributing particles while exactly conserving total intensity per color channel. }
    \label{fig:schematic}
    \vspace{-3mm}
\end{figure*}

Generative diffusion modeling based on discrete-state Markov chains has become an active area of research in recent years. Early work, such as \citet{hoogeboom2021argmax,austin21}, introduced discrete-state and discrete-time Markov chains as an alternative to the Gaussian noise used in conventional diffusion models \citep{sohl-dicksteinDeepUnsupervisedLearning2015,hoDenoisingDiffusionProbabilistic2020,songScoreBasedGenerativeModeling2021a}. \citet{campbell2022} generalized these formulations to a continuous-time framework, providing a more rigorous theoretical foundation for discrete-state generative diffusion modeling. \citet{blackout} employed operator algebraic analysis to formally establish the existence of the reverse-time dynamics and derived the stochastic generator for arbitrary discrete-state Markov processes. Similar formulations were independently developed by \citet{sunScorebasedContinuoustimeDiscrete2022} and \citet{louDiscreteDiffusionLanguage2023}, with an emphasis on defining and estimating score functions for discrete-state systems. The Markov process operates in intensity space in all the aforementioned diffusion models, treating each pixel as an independent stochastic process (Fig.~\ref{fig:schematic}(a): Gaussian; Fig.~\ref{fig:schematic}(b): Discrete). 

This study focuses on a spatially correlated process for generative modeling for two reasons: (1) for structured images, it is more natural to incorporate spatial correlations into the generative process, and (2) spatially decorrelated noise makes it difficult to preserve total intensity. A spatially correlated approach has been explored for continuous systems. Cold Diffusion \citep{Bansal2022ColdNoise} introduced a deterministic blurring transformation, where image degradation follows a predefined forward process, and reconstruction is learned as an inverse mapping. However, lacking a probabilistic latent distribution (as in VAEs \citep{kingmaAutoencodingVariationalBayes2014}), Cold Diffusion is not a true generative model. Inverse Heat Dissipation Model (IHDM, \citet{heat_eq}) uses the heat equation as a corruption model. Since the heat equation is deterministic and reversible (except that the homogeneous solution at $t\rightarrow \infty$ is singular), a na\"ive inversion would again result in deterministic reconstructions. Uncorrelated Gaussian noise was added to the heat equation to overcome this limitation, relaxing the deterministic process into a probabilistic It\^o diffusion. Later, Blurring Diffusion Model (BDM, \citet{hoogeboomBlurringDiffusionModels2022}) recognized that IHDM could be recast as a Gaussian diffusion model in the spectral domain. BDA extended IHDM and achieved SOTA generative performance, validating the hypothesis that spatially structured diffusion processes can enhance image generation. Nevertheless, the probabilistic formulation of IHD and BDM only preserves intensity on average, not exactly per-sample, and their continuous-state nature makes it difficult to apply to discrete datasets. 

Our goal of generating samples with exactly conditioned total intensity aligns with conditional diffusion modeling. However, existing approaches all rely on some degree of approximation. \citet{songGenerativeModelingEstimating2020} proposed a simple conditional sampling method by passing class labels into the neural network during training, but this does not guarantee exact enforcement of the condition in generated samples. A more structured approach was introduced by \citet{chungScorebasedDiffusionModels2022,chungComeCloserDiffuseFasterAcceleratingConditional2022,chungImprovingDiffusionModels2022}, which interleaved projection steps with diffusion sampling to enforce linear constraints in image generation. However, these projections disrupt the exactness of the forward corruption and reverse inference dynamics \citep{andersonReversetimeDiffusionEquation1982,campbell2022,blackout}, leading to a mismatch between the projected and true data manifolds. To address this, \citet{chungDiffusionPosteriorSampling2022} eliminated projection steps but instead relaxed deterministic constraints into a probabilistic formulation via a noisy measurement model. However, this method does not apply to deterministic constraints, as it becomes singular in the limit of zero measurement noise. An alternative approach leverages Bayes’ theorem for a posteriori conditional sampling, that is, $ p(\text{S} \vert \text{C}) \propto p(\text{S}) p(\text{C} \vert \text{S})$, where ``S'' stands for samples and ``C'' for condition(s). Because $p(\text{S})$ is given by a trained unconditional diffusion model, conditioning can be performed if one has $p(\text{C} \vert \text{S})$, which is however intractable for arbitrary data distributions\footnote{It is challenging because the constraint is imposed on the \textit{final samples at the end of the inference}, but the conditioning ``S'' are \textit{samples generated during the inference}.}. Existing methods approximate this term crudely or by training a separate classifier as in \citet{songScoreBasedGenerativeModeling2021a}, or by a Gaussian approximation with moment-matching as in \citep{finziUserdefinedEventSampling2023a,duCoNFiLDConditionalNeural2024}. None of these methods guarantees that the generated samples are exactly conditioned.

The principal contribution of our work is that it provides a new capability for diffusion models to preserve intensity exactly in a fully discrete-state context. The approach is based entirely on how the diffusion process is built, and how the model is trained; it is readily usable with existing diffusion model neural network (NN) architectures. To our knowledge, this is the first diffusion model to incorporate  spatially correlated noise, achieved through a stochastic jump process that allows units of intensity to perform a random walk. This fact also demonstrates that more complex noise processes can themselves be tractable. We furthermore demonstrate that such a model is effective for conventional image synthesis tasks. The relevance and power of the approach is then demonstrated through application to scientific data in the field of materials microstructure, where the ability to generate complex data-driven images with constrained total intensity is key to physically meaningful generative modeling and scientific reliability.

\section{Methods}

\subsection{Corruption Process}

In this manuscript, we adopt the language of image processing and consider 2-dimensional images, although the application context and spatial dimensionality of the data are not constrained by the mathematical framework provided here. We treat a digital image with discretized intensity values $I_{x,y,c}\in \mathbb{Z}_{\ge 0}$ at pixel $(x,y)\in \{1, \ldots, W\} \times \{1, \ldots, H\}$ in color channel $c \in \{1, \ldots, C\}$. Within the DSD framework, the intensity values in the image are treated as a spatially organized collection of particles, with one particle for each intensity unit. Below, we will interchangeably use ``particles'' and ``intensity units'' to denote these fundamentally discrete units. Specifically, $I_{x, y, c} = n$ implies $n$ particles of type $c$ at location (x,y), and the total number of particles of the system is $\sum_{x=1}^H \sum_{y=1}^W \sum_{c=1}^C I_{x,y,c}$. In the forward stochastic process with the time parameter $t$, each of the particles in the system \textit{independently} performs a continuous-time and discrete-state random walk:
\begin{subequations} \label{eq:forwardProcess}
\begin{align}
    (x,y,c) \xrightarrow{r}{}&   (x + \nu_x, y+\nu_y,c), \\
    \nu  := \left(\nu_x, \nu_y\right) \in{}&   \left\{\left(1,0\right), \left(-1,0\right),\left(0,1\right), \left(0,-1\right)\right\}
\end{align}
\end{subequations}
where $r$ is the transition rate of the particle jumping to one of their nearest neighbors, and $\nu$ is a set of four directions the particles can hop to their nearest neighbors. A schematic diagram is shown in Fig.~\ref{fig:schematic}c. Note that the particles perform jumps in the $(x,y)$ space at random times, but do not change their color coordinate $c$. Because of this, the forward process conserves the total number of particles $\sum_{x=1}^H \sum_{y=1}^W I_{x,y,z}$ in each color channel independently.  We impose either no-flux boundary condition, such that the transition rate for jumps out of the image domain are zero, or periodic boundary conditions, so that a jump to $x=W+1$ becomes a jump to $x=0$, vice-versa, and analogously for $y$.

\begin{figure*}[ht!]
    \centering
    \includegraphics[width=1\linewidth]{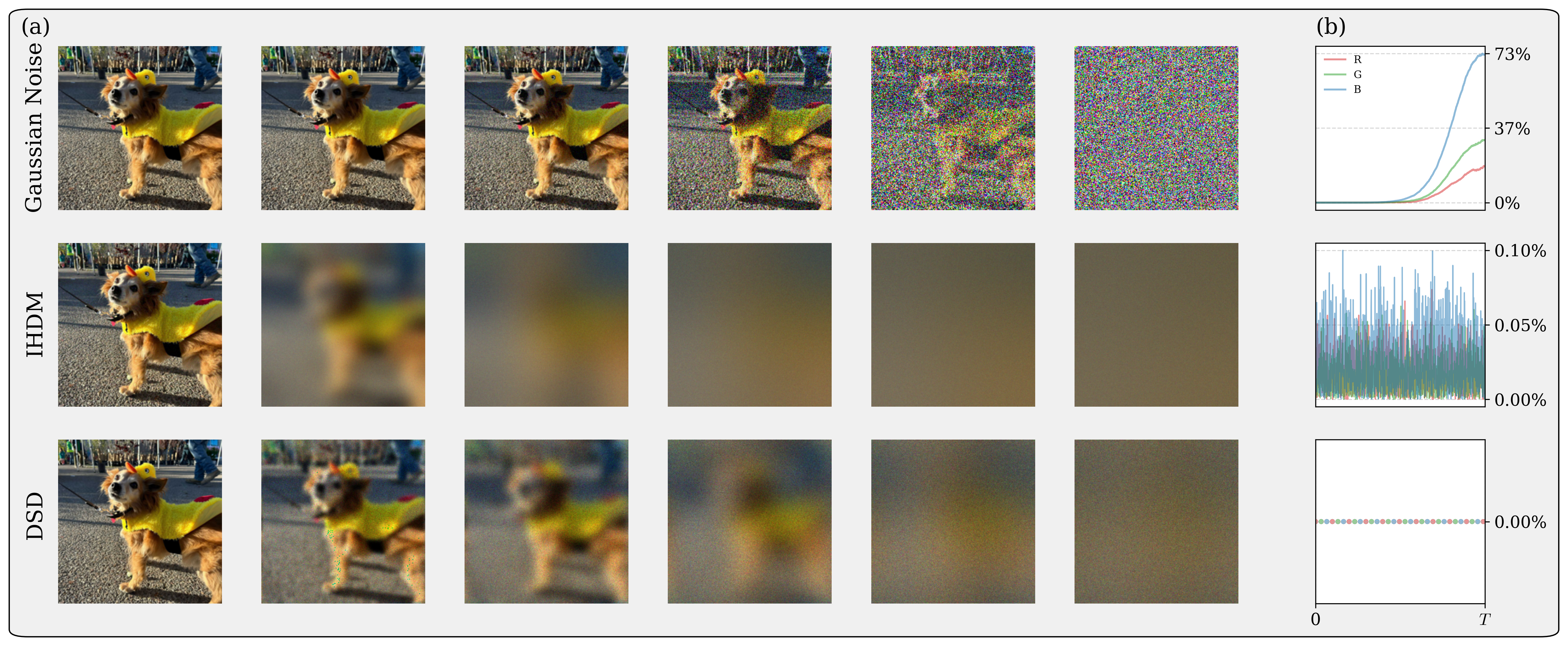}
    \vspace{-7mm}
    \caption{\textbf{(a)} The forward processes for Gaussian Diffusion \cite{hoDenoisingDiffusionProbabilistic2020}, Inverse Heat Dissipation Model \cite{heat_eq}, and Discrete Spatial Diffusion (ours) applied on an image, sampled at discrete times. \textbf{(b)} Percentage change in intensity relative to the original image under the forward process.}
    \label{fig:processes}
\end{figure*}

We refer to the spatial hopping process (Eq.~\eqref{eq:forwardProcess}) as the \textit{Discrete Spatial Diffusion} (DSD), noting the ``discreteness'' refers to both the discretized intensity units and the discreteness of the spatial lattice $\{1, \ldots, W\} \times \{1, \ldots, H\}$ where the particles are allowed to reside. DSD, as well as similar discrete-state random walks, have been extensively studied in non-equilibrium statistical physics and stochastic processes (\citet{vanKampen}, \citet{gardinerStochasticMethodsHandbook2009}, \citet{giuggioliExactSpatiotemporalDynamics2020} and references therein). The evolution of the probability distribution of the single random walk in the continuum space limit, under the appropriate scaling of the transition rate \cite{einsteinUeberMolekularkinetischenTheorie1905}, converges to the Fokker--Planck Equation (FPE, \citet{vanKampen,riskenFokkerPlanckEquation1984}), which is mathematically identical to the heat equation. Because of the duality between the probabilistic FPE and the deterministic heat equation \citep{lawlerRandomWalkHeat2010}, DSD can be considered as a microscopic description of the macroscopic heat dissipation that inspires IHD and BDM. Notably, the correlated noise is built in DSD, in contrast to the heuristic addition of uncorrelated Gaussian noise in IHD and BDM. Fig.~\ref{fig:processes} illustrates the application of DSD to a sample image. Due to the stochasticity of the random jumps, the limiting behavior ($t\rightarrow \infty$) of this process is a random configuration with no discernible structure or similarity to the original spatial organization aside from the conserved global particle counts in each color channel.

We use $(X_t, Y_t,C_t)$ to denote the random process in $(x,y,c)$ space, and $(x_0, y_0, c_0)$ are the initial condition of a specific particle. We use $I_t$ to denote the randomly corrupted image at the time $t$, where $\left[I_t\right]_{x,y,c}$ is the total number of particles at $(x,y)$ in color channel $c$. The process can be represented in these two dual representations: with $(X_t, Y_t, C_t)$ the process is formulated in the frame of a moving particle (the Lagrangian frame), and with $I_t$ the process is formulated as a histogram in space-time (the Eulerian frame). Below, we will use these two representations interchangeably. 

The forward solution and the transition probabilities $p_t(x,y,c\vert x_0,y_0,c_0):= \mathbb{P}\{X_t =x, Y_t=y, Z_t = z\vert X_0 = x_0, Y_0=c_0, C_0=c_0\}$, can be computed by integrating the Master Equation \citep{vanKampen,gardinerStochasticMethodsHandbook2009,weberMasterEquationsTheory2017}. This corresponds to exponentiating the Markov transition matrix of the process defined in Eq.~\eqref{eq:forwardProcess}. While the matrix exponential required numerically for no-flux boundaries is expensive, the solution can be stored and reused to corrupt images and to compute the reverse-transition rates (see Sec.~\ref{sec:reverse}) for learning. When periodic boundary conditions are imposed, the transition matrix is diagonal in the discrete Fourier space, facilitating the efficient computation of $p_t(\cdot\vert \cdot)$ (see Sec.~\ref{app:DSD-FT}). 

\subsection{Designing Noise Schedules by Structural Similarity Index Metric (SSIM)}

Since the corruption process is time-homogeneous \eqref{eq:forwardProcess}, the noise applied to each particle remains constant over time. However, it has been shown that inhomogeneous noise schedules can facilitate learning \citep{nicholImprovedDenoisingDiffusion2021}. We use the formulation of a recent study \cite{santosUnderstandingDenoisingDiffusion2023} identified the unique correspondence between non-uniform observation times in a homogeneous Ornstein--Uhlenbeck process \citep{OU} and noise schedule in conventional diffusion models \cite{hoDenoisingDiffusionProbabilistic2020,song2020denoising}. We follow the same philosophy as \citet{santosUnderstandingDenoisingDiffusion2023} to construct a sequence of observation times $t_0=0 < t_1 < t_2 < \ldots < t_T =1$, at which we will generate random samples for learning. Here, $T$ denotes the total number of discrete time steps used to generate corrupted sample images during training.

We adopt a heuristic approach to construct the discrete times. The idea is to use a metric to quantify how much the ``quality'' of the images has been degraded up to time $t$, and we aim to design $t_k$'s such that the metric degrades
from $k=0$ to $k=T$ as evenly as possible. We chose the Structural Similarity Index Metric (SSIM, \citet{SSIM}) between the corrupted image and the original image. We generalize a generic monotonic relation between $k$ to $t_k$ proposed by \citet{blackout}:
\begin{align}
    \Phi \left(e^{-\tau_2 t_k}\right) \triangleq \frac{\left(k-1\right)\Phi\left(e^{-\tau_2}\right) - \left(T-k\right)\Phi\left(e^{-\tau_1}\right) }{T-1},\label{eq:observationTimes}
\end{align}
where $\Phi(p):= \log p/(1-p)$ is the logit function, $\tau_1$ and $\tau_2$ are parameters used to construct the observation times. Note that $t_T=1$ in the above parametrization. Specifically, we tune $\tau_1$, $\tau_2$ and the unit transition rate $r$ in process \eqref{eq:forwardProcess}, using a subset of training samples, aiming to cover an even degradation of the SSIM throughout observation times. We found that setting $\tau_1=7.5$ and $\tau_2=2.5$, and $r=120$-$160$ is sufficient for numerical experiments. We remark that the choice of the functional form in Eq.~\eqref{eq:observationTimes} is arbitrary and without any theoretical foundation; we only treat Eq.~\eqref{eq:observationTimes} as a versatile monotonic fitting function, whose corresponding SSIM degradation is empirically more symmetric than polynomial and cosine schedules \cite{nicholImprovedDenoisingDiffusion2021} for the DSD process (see Appendix Fig.~\ref{fig:SSIM}). 

\subsection{Reverse-time process}\label{sec:reverse}
Following the process formalism from \citep{campbell2022,blackout}, there exists a reverse-time process that evolves in opposite time and whose joint probability distribution is identical to that of the forward process \eqref{eq:forwardProcess}. Specifically, the reverse-time process corresponding to process \eqref{eq:forwardProcess} is:
\begin{equation}
    (x,y,c) \xrightarrow{r \frac{p_t(x +\bar{\nu}_x,y +\bar{\nu}_y,c\vert x_0,y_0,c_0)}{p_t(x,y,c\vert x_0,y_0,c_0)}} (x + \bar{\nu}_x, y+\bar{\nu}_v,c), \label{eq:reverseProcess}
\end{equation}
where the admissible reverse-time transitions $\bar{\nu}=(\bar{\nu}_x,\bar{\nu}_y):=\in \left\{\left(-1,0\right), \left(1,0\right),\left(0,-1\right), \left(0,1\right)\right\}$ are the reversed direction of the forward jumps ($\bar{\nu}_x=-{\nu}_x, \, \bar{\nu}_y=-{\nu}_y$).  
The framework ensures the same boundary condition to be imposed (no-flux or periodic, according to the forward process). We note that the reverse-time process, and therefore the generated images, also conserve the total particle number per color channel.

We note that the reverse transition rate depends on both the initial condition $(x_0,y_0,c_0)$ of a particle and the forward solution $p_t\left(x,y,c\vert x_0, y_0, c_0\right)$, $\forall (x,y,c)$. This is analogous to conventional diffusion models, where either the reverse-time drift \cite{sohl-dicksteinDeepUnsupervisedLearning2015,hoDenoisingDiffusionProbabilistic2020} or the score function \citep{songScoreBasedGenerativeModeling2021a} formally depend on the initial sample and the solution of the forward process. However, during the inference, the initial particle configuration is not known, and as such, we train an NN to learn the reverse transition rates using samples $I_t$ generated from the forward process \eqref{eq:forwardProcess} at $t>0$. Additionally, the particles are indistinguishable, but the rate prescribed in Eq.~\ref{eq:reverseProcess} is \emph{per-particle}, raising the question: what is the appropriate \emph{per-pixel} reverse transition rate that the NN ought to model? This question can be answered by performing the survival analysis of the many-particle system in light of the independence of particle motion; see Appendix \ref{app:reverseRate} for a derivation. The analysis shows that the reverse transition rate of the first jump of $n=\left[I_t\right]_{x,y,c}$ particles is simply the sum of the instantaneous transition rates:
\begin{equation}
    \bar{r}_{\bar{\nu},x,y,c} = r \sum_{i=1}^n \frac{p_t(x + \bar{\nu}_x,y + \bar{\nu}_y,c\vert x_0^{[i]},y_0^{[i]},c_0^{[i]})}{p_t(x,y,c\vert x_0^{[i]},y_0^{[i]},c_0^{[i]})}, \label{eq:reverseRates}
\end{equation}
Eq.~\ref{eq:reverseRates} prescribes the rate for \textit{the first} of all the particles (which is $\left[I_t\right]_{x,y,c}$) to jump to one of its neighboring pixels. Intuitively, this can also be derived by combining the first-reaction method \citep{gillespieGeneralMethodNumerically1976a} and inhomogeneous Poisson process (e.g., see \citet{corbellaAutomaticZigZagSampling2022a}). Eq.~\ref{eq:reverseRates} also prescribes the rate that the NN will model. This rate is implicitly time-dependent through the dependence on the forward solution $p_t$, similar to standard continuous-time diffusion models. 

\subsection{Loss functions}
Our goal is to provide the corrupted images $I_t$ at a sampled time $t > 0$ to a neural network (NN) and to train it to predict the reverse-time transition rates $\bar{r} \in \mathbb{R}_{+}^{4 \times H\times W\times C}$ of Eq.~\ref{eq:reverseRates}. The first dimension (of size four) represents rates for the four nearest-neighbor transitions. We denote the NN modeled rates as $\bar{r}^\text{NN} (I_t, t)$ with noisified image $I_t$ at time $t$. 

We use two approaches to formulate the loss function. The first and more commonly used approach adopts a metric to heuristically match the NN prediction and the ground truth, as in DDPM~\cite{hoDenoisingDiffusionProbabilistic2020}, score-matching~\cite{songScoreBasedGenerativeModeling2021a}, and flow-matching~\cite{lipmanFlowMatchingGenerative2022}.  
We extend these schemes to \emph{rate-matching}, where we minimize a chosen norm of the difference between the predicted and true rates $\bar{r}^\textrm{NN}$ and $\bar{r}$. For example, for using $\text{L1}$, a loss $\mathcal{L}$: 
\begin{equation}
\mathcal{L}_\text{rate-matching} = \mathbb{E}_{I_{t_k}, k}  \left[\text{mean}( \left \vert \bar{r}^\text{NN} -  \bar{r}\right \vert) \right]. \label{eq:l1loss}
\end{equation}
Here, $k \in \left\{1, \ldots T \right\}$ is uniformly sampled, $I_{t_k}$ is drawn from the random process \eqref{eq:forwardProcess} at the sampled times, $\bar{r}=\bar{r}_{\bar{\nu}, x,y,c}(I_t, t\vert I_0 )$ is the true reverse-time transition rate \eqref{eq:reverseRates}, $\bar{r}^\text{NN}=\bar{r}^\text{NN}_{\bar{\nu}, x,y,c}(I_t, t)$ is the NN-predicted reverse-time transition rate, and the mean is over all the indices $(\bar{\nu}, x,y,c)$.
The second and more principled approach is through minimization of the negative log-likelihood $L$ of the NN-induced process to predict the analytical reverse-time process \cite{sohl-dicksteinDeepUnsupervisedLearning2015,campbell2022,blackout}:
\begin{equation}
     \mathcal{L}_\text{likelihood} = - \log L = - \mathbb{E}_{I_t}  \left[\int_0^\infty \sum \left( \bar{r}^\text{NN} -  \bar{r}\log \bar{r}^\text{NN}  \right) {\rm d} t\right]. \label{eq:likelihood-cts}
\end{equation}
Because we only observe the process at discrete times prescribed in Eq.~\eqref{eq:observationTimes}, we approximate the continuous-time integration above by 
\begin{equation}
     \log L =\mathbb{E}_{I_{t_k}, k}  \left[ \left(t_k - t_{k-1}\right) \sum \left( \bar{r}^\text{NN} -  \bar{r}\log \bar{r}^\text{NN}  \right)\right], \label{eq:likelihood}
\end{equation}
where we again take expectation over randomly sampled $t_k$ and $I_k$. 
In this study, we experimented with both loss functions and observed no  noticeable difference, giving evidence that the DSD forward process \eqref{eq:forwardProcess} is not sensitive to the choice of the loss function. This stands in contrast to Gaussian diffusion models, where training loss choice significantly impacts performance  \citet{hoDenoisingDiffusionProbabilistic2020}, which used the heuristic approach to improve over \citet{sohl-dicksteinDeepUnsupervisedLearning2015}, which adopted the second approach. We focus on learning the transition rates of the reverse-time dynamics, which is distinct from the ratio-matching approach \cite{sunScorebasedContinuoustimeDiscrete2022,louDiscreteDiffusionLanguage2023} which focuses on learning the probability distribution $p_t(\cdot)$, although a similar formulation (``implicit score entropy'') proposed by \cite{louDiscreteDiffusionLanguage2023} can be regarded as the process likelihood \eqref{eq:likelihood} first proposed in \citet{blackout}.  Algorithm \ref{alg:training} describes the DSD training pseudocode.

\subsection{Sampling with an Adaptive Time Stepping using the Courant--Friedrichs--Lewy condition}
Once trained, the neural network will predict reverse-time rates \eqref{eq:reverseRates} from the given the configuration of system $I_t$, at time $t\ge 0$. Because the reverse rates are time-dependent, one could generate the exact sample paths of the inhomogeneous Poisson process by integrating the survival function of the first reaction on each pixel in each color channel (see e.g., algorithms reported in \citet{corbellaAutomaticZigZagSampling2022a}). However, this approach is not computationally efficient, so we resort to $\tau$-leaping \cite{tauLeaping21}, an integrator that has been adopted by essentially all continuous-time and discrete-state diffusion models~\cite{campbell2022,blackout,bridgingDiscreteContinuous24,renHowDiscreteContinuous2024}, analogous to the Euler's method for ordinary or partial differential equations and Euler--Maruyama for It\^o SDEs. The central idea of $\tau$-leaping is to approximate the reverse-time transition rates $\bar{r}$ as a fixed constant in a small enough window $(t-\tau, t)$, assuming the time-dependent rates change slowly in the period, a condition often termed as the ``leap condition'' \cite{tauLeaping21,caoAvoidingNegativePopulations2005}. With this assumption, the original $\tau$-leaping algorithm by \citet{tauLeaping21} generates Poisson random numbers to update the system's discrete states. However, this approach could sometimes lead to a negative population of particles, which cannot happen in the process, due to violations of the leap condition. Mitigation strategies exist \cite{gillespieImprovedLeapsizeSelection2003,caoAvoidingNegativePopulations2005,caoEfficientStepSize2006}, however, some of them are limited to small reaction networks and not suitable for the DSD sampling task, which involves a very large number of ($4 \times H \times W \times C$) of transition rates to estimate.

As such, we propose a more efficient (but arguably cruder) approach to select the step size $\tau$ adaptively. Our idea is to combine the binomial $\tau$-leaping \cite{tianBinomialLeapMethods2004, chatterjeeBinomialDistributionBased2005} and the Courant--Friedrichs--Lewy (CFL) condition \cite{courantUberPartiellenDifferenzengleichungen1928} to conservatively determine the adaptive step size $\tau$. Specifically, since the jump scale is fixed at the pixel length scale, the timescale $\tau$ fully determines the CFL condition. The idea is to choose a $\tau$ such that the CFL number is fixed throughout the inference\footnote{Even though CFL condition is more commonly used in PDE integrators, the concept can be applied for our stochastic system. Suppose the reverse-time rate is $\bar{r}$. On average, the particle would move at a timescale $1/\bar{r}$ to one of its neighbors, traveling $\Delta x$. Then, the velocity $c=r \Delta x$. The CFL condition is then $c \Delta t / \Delta x$ where in our scheme $\Delta t$ is the $\tau$; thus, the classical CFL convergence condition translates to the obvious bound of transition probability $\bar{r} \tau <1$. This motivates us to ensure a conservative estimation of $\tau$, but enforcing a small $\bar{r} \tau$ to reduce the error.}. To achieve this, we compute the reverse-time transition rates $\bar{r}_{\bar{\nu}}$ for each pixel in each channel, noting that the probability of a particle in that channel will jump to one of its neighboring locations is $\bar{r}_{\bar{\nu}} \tau$. Then, we determine $\tau$ by fixing the largest probability across all the pixels and color channels at a constant. Algorithm \ref{alg:inference} describes the DSD inference pseudocode.

\section{Computational Experiments}

We employ the Noise Conditional Score Network (NCSN++) \cite{songScoreBasedGenerativeModeling2021a,alexandreadam_score_models} with two modifications: the final convolutional layer outputs 4 times the number of input channels (e.g., 3 for RGB) to represent four directions (up, down, left, right), and we use a SoftPlus activation function to ensure non-negativity in the predicted rates. The hyperparameters used can be found in Appendix \ref{app:hparams}.

\subsection{Image synthesis benchmarks}

While Discrete Spatial Diffusion (DSD) was developed to enable generative modeling under a strict intensity constraint, we first evaluate its performance on conventional image synthesis benchmarks. Specifically, we trained DSD models on MNIST \cite{mnist}, CIFAR-10 \cite{krizhevsky2009learning}, and CelebA \cite{celebA}, suggesting that DSD can approximate key statistical and structural properties of image distributions, despite operating under discrete-state and global conservation constraints.

\begin{figure*}[!ht]
\vspace{-4mm}
    \centering
    \includegraphics[width=1.\linewidth]{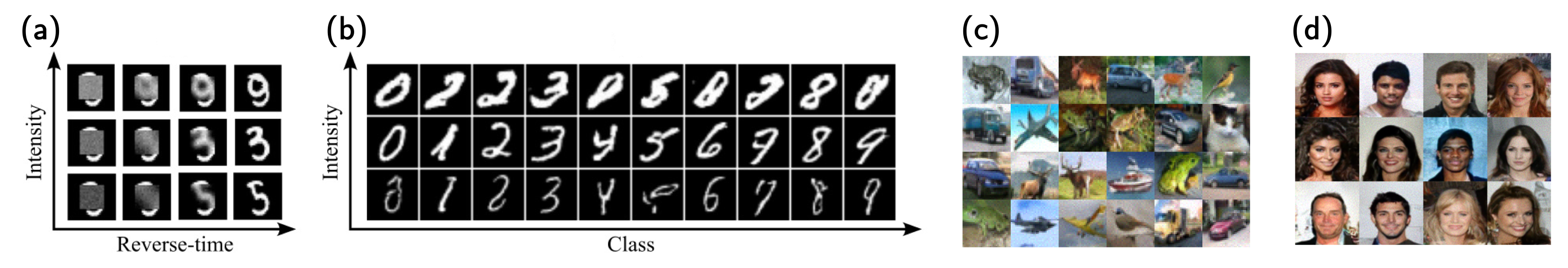}
    \vspace{-7mm}
    \caption{ \textbf{(a)} Inpainting realizations on MNIST; 15\% difference of conditioning intensity between consecutive rows. \textbf{(b)} Conditioned MNIST generations across different intensities and classes.   \textbf{(c)} Unconditional CIFAR-10 generations. \textbf{(d)} Unconditional CelebA generations.  }
    \label{fig:image_applications}
\end{figure*}

We begin with MNIST, where intensity naturally corresponds to stroke thickness and digit area.  In Fig.~\ref{fig:image_applications}(a), we show inpainting results with a fixed mask with no-flux boundary conditions. During inference, particles inside the masked region rearrange to complete the digit structure, while the surrounding region remains fixed. Holding the visible structure constant, we varied the number of particles allowed to move, revealing that the total intensity governs which digit is most likely to emerge. This highlights the ability of DSD to incorporate hard constraints in downstream tasks such as inpainting. Additionally, we trained a conditional DSD model that employed a standard class-conditioning \cite{songScoreBasedGenerativeModeling2021a}. In Figure~\ref{fig:image_applications}(b) we illustrate the class-conditioned generated images with different total numbers of particles, varying from low, typical, to high total intensities. While these do exhibit some artifacts, DSD notably learns the spatial structure of the digits and generates ``Bolder'' or ``Lighter'' digits without saturating the upper bound of the intensity (i.e. 255 for \texttt{uint8}). This would not have been precisely realizable using conventional diffusion models. Comparisons with other conditioning approaches can be found in Appendix~\ref{sec:comparison}.

We also evaluated DSD on RGB datasets to explore scalability and generality. In Fig. \ref{fig:image_applications}(c,d), we show unconditional generations from models trained on CIFAR-10 and CelebA. Despite the discrete state space and intensity-preserving dynamics, the model captures complex semantic structures from lighting, and textures to  animals, vehicles, and human facial features. Appendix~\ref{sec:human-centric} includes large grids of generated samples, Fréchet Inception Distance (FID), and spatial FID (sFID) metrics, sampling ablations, and post-processing strategies for improving sample smoothness.

\subsection{Subsurface rock microstructures}

The microstructure of subsurface rocks governs a wide range of physical processes, including fluid transport, electrical resistivity, and mechanical deformation \cite{blunt2013}. This originates from connected pores on the nano- and micro-scale, which vary in size, structure, and coordination degree across rock types. High-resolution 3D imaging via X-ray microtomography enables detailed pore-scale reconstructions, but these scans are expensive and limited to sample sizes on the order of millimeters to centimeters \cite{cnudde2013high}. While direct imaging of rock microstructure is costly, measuring porosity (defined here as average intensity over the image) across large formations is inexpensive and can be performed without specialized equipment \cite{simplifiedporositymeasurements,practicalporosity}. This enables large-scale field measurements of porosity, even when high-resolution microstructural data is unavailable.

\begin{figure*}[h!]
\vspace{-4.5mm}
   \centering
    \includegraphics[width=1.0\linewidth]{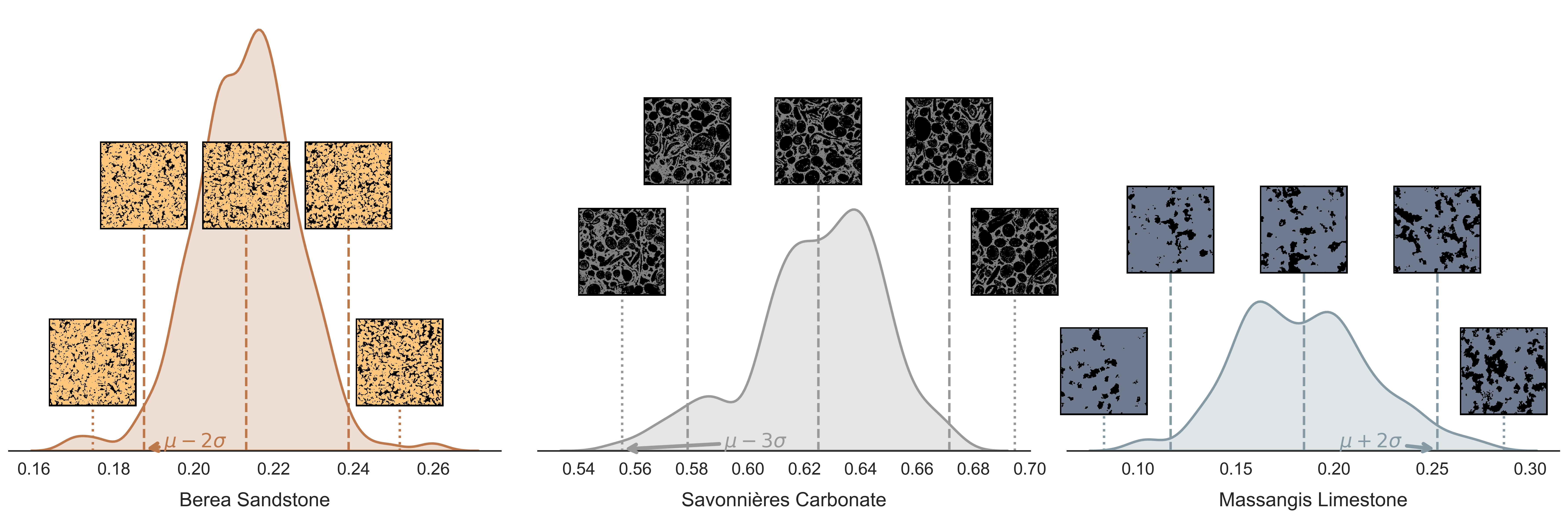}
\vspace{-6.75mm}
    \caption{Distributions of porosity and generated samples for three rock classes. Each plot shows the porosity distribution across the dataset (shaded curve) and corresponding DSD-generated microstructures at selected quantiles. \textbf{(Left)} Berea Sandstone, \textbf{(center)} Savonnières Carbonate, and \textbf{(right)} Massangis Limestone. Samples illustrate the model’s ability to generate structurally diverse images conditioned on exact global intensity. Quantitative metrics are presented in Appendix~\ref{app:geology_metrics}.}
    \label{fig:rocks}
 \end{figure*}


To overcome this limitation, synthetic models are frequently used to generate representative pore structures for computational physics studies \cite{processbased}. However, conventional reconstruction techniques impose strong geometric assumptions that fail to capture the heterogeneity observed in real rocks like Berea Sandstone, Savonnières and Massangis Carbonates. We trained DSD models using these rock samples, which represent a broad spectrum of pore structures (including granular, fossiliferous, and dissolution-driven features) across two lithologies. A  description of the training datasets is provided in Appendix~\ref{app:geology}.  Fig.~\ref{fig:rocks} presents representative outputs from our models trained on 256$\times$256 binary images. The generated samples accurately reproduce the key morphological and statistical features of the original datasets, including two-point spatial correlations and pore size distributions which are primary microstructural descriptors that govern permeability, tortuosity, and diffusion behavior in porous media (Fig.~\ref{fig:pore_metrics}). Given that DSD allows for precise control over total porosity, one can generate synthetic microstructures that match the porosity measured in the field, enabling the reconstruction of representative pore-scale samples even in the absence of direct imaging.

\subsection{Lithium-ion electrodes}

Electrodes in lithium-ion batteries are porous materials with a complex microstructure that governs key properties like ion transport and electrochemical performance.
Nickel-manganese-cobalt cathodes, among the most common, are composed of three phases: the active material driving the electrochemical reaction, the carbon binder ensuring electrical conductivity and mechanical stability, and the pore space filled with electrolytes. 
\begin{wrapfigure}{r}{0.5\linewidth}
    \vspace{-1.6em}
    \centering
    \includegraphics[width=\linewidth]{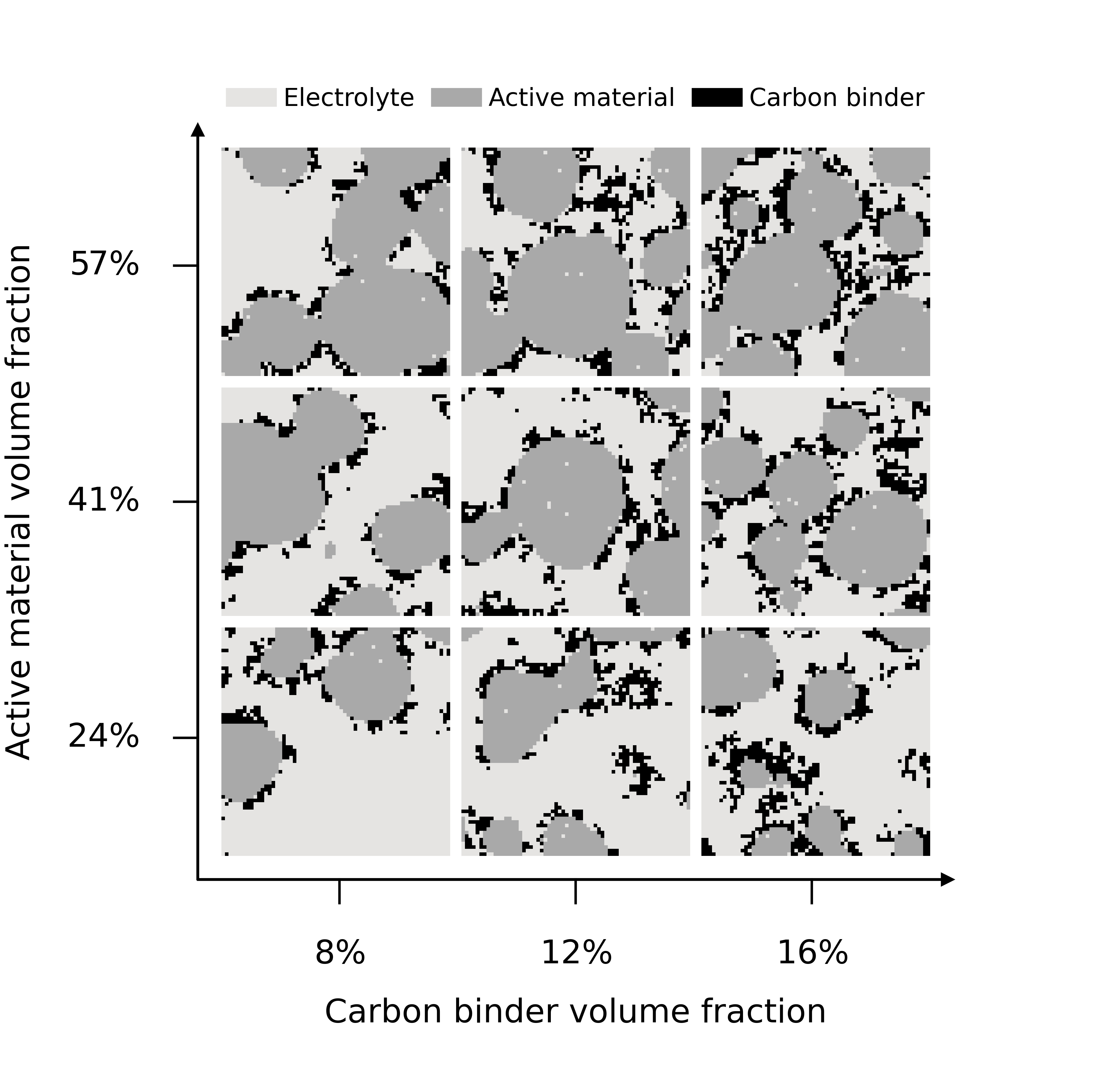}
    \vspace{-2.6em}
    \caption{Generated cathode microstructures with varying phase volume fractions. The carbon binder domain appears in black, active material particles in gray, and electrolyte-filled pore space in white.}
    \label{fig:electrode_samples}
\end{wrapfigure}
The active material is expensive, creating a strong economic incentive to understand how its volume fraction and distribution influence electrode behavior.
While tomographies are needed for studying microstructures and enabling computational modeling, acquiring diverse datasets is challenging~\cite{deng2021recent}. To overcome this, researchers often rely on computational methods to generate synthetic microstructures~\cite{duquesnoy2023machine}. While GANs have been explored for this purpose, they do not control phase volume ratio~\cite{gayon2020pores}. We trained a DSD model on tomography data~\cite{usseglio2018resolving}, where two color channels were used to represent the carbon binder and active materials.  The results, shown in Fig.~\ref{fig:electrode_samples}, demonstrate DSD enables precise tuning of phase volume fractions. We then computed key morphological metrics characterizing the electrode structure~\cite{kench2023taufactor}, demonstrating the strong generative capability of the DSD method, which generalizes across the training set more effectively than existing approaches in the literature. For more details on datasets and reconstruction metrics applied to these samples, see Appendix~\ref{app:cathodes}.


\section{Limitations}
\label{sec:limitations}
The computational cost of forward sampling during training and reverse-time sampling during inference in DSD scales linearly with the \emph{total intensity} of the image. While this makes DSD highly efficient for low-bit-depth or binary datasets, it may become less efficient than other techniques for higher-resolution images or datasets with higher intensity saturation, such as standard \textit{uint8} images. While the FLOP cost of the forward noising process is higher than in conventional Gaussian diffusion models, this is handled entirely by CPU workers during the data loading phase, and so the added cost is effectively hidden completely behind the data pipeline and does not impact GPU throughput; DSD models were negligibly slower than their Gaussian counterparts, and models were trained in this work using one A100 GPU in $<\!750$ hours per model (RGB images) and $<\!50$ hours per model (microstructures, MNIST) . Inference is slower (approximately 2$\times$ per), due to the need to sample individual particle transitions, but affordable across datasets of practical interest. Despite this cost scaling, many of the largest available microstructural datasets in subsurface modeling are on the order of $1000 \times 1000$ resolution. We trained DSD on such large-scale scientific data, demonstrating that the method remains computationally feasible in this regime, also suggesting the feasibility of  extending it to 3D by carrying out implementation changes. Representative results from high-resolution generation are presented in Appendix~\ref{sec:large_rock}. Additionally, enforcing strict intensity conservation requires a custom forward process code (Eq.~\eqref{eq:forwardProcess}) and a novel sampling scheme, deviating from conventional Gaussian diffusion models. This introduces a steeper learning curve for practitioners accustomed to standard diffusion approaches. However, we argue that these trade-offs are necessary to achieve exact intensity constraints, which is not possible with existing methods.

\section{Conclusion}

We introduced Discrete Spatial Diffusion (DSD), a fully discrete, intensity-preserving generative model approach for images and scientific data. The foundation is the use of discrete-state, continuous time statistical processes incorporating jump dynamics, rather than SDEs, and in particular is the first discrete diffusion model to explore spatially correlated noisification. DSD demonstrates competitive quality on standard benchmarks while enabling exact global constraints on particle count (thereby conserving image intensity or mass fractions under various applications) that are critical in many scientific applications. By preserving these constraints in both forward and reverse processes, DSD enables exactly constrained data generation, which we explored on image synthesis and domain-specific datasets. It also demonstrates that more complex statistical processes (in this case, random walks) can be used for diffusion modeling, opening the door for further models to exploit structure in their dynamics such as conservation laws and symmetries.

\bibliography{refs}

\begin{thebibliography}{80}
\providecommand{\natexlab}[1]{#1}
\providecommand{\url}[1]{\texttt{#1}}
\expandafter\ifx\csname urlstyle\endcsname\relax
  \providecommand{\doi}[1]{doi: #1}\else
  \providecommand{\doi}{doi: \begingroup \urlstyle{rm}\Url}\fi

\bibitem[Adam(2023)]{alexandreadam_score_models}
Alexandre Adam.
\newblock score\_models: A repository to store state of the art score model
  architectures.
\newblock \url{https://github.com/AlexandreAdam/score_models}, 2023.
\newblock Accessed: 2025-05-13.

\bibitem[Anderson(1982)]{andersonReversetimeDiffusionEquation1982}
Brian~D.O. Anderson.
\newblock Reverse-time diffusion equation models.
\newblock \emph{Stochastic Processes and their Applications}, 12\penalty0
  (3):\penalty0 313--326, 1982.
\newblock ISSN 0304-4149.
\newblock \doi{https://doi.org/10.1016/0304-4149(82)90051-5}.

\bibitem[Austin et~al.(2021)Austin, Johnson, Ho, Tarlow, and van~den
  Berg]{austin21}
Jacob Austin, Daniel~D. Johnson, Jonathan Ho, Daniel Tarlow, and Rianne van~den
  Berg.
\newblock Structured denoising diffusion models in discrete state-spaces.
\newblock In Marc'Aurelio Ranzato, Alina Beygelzimer, Yann~N. Dauphin, Percy
  Liang, and Jennifer~Wortman Vaughan, editors, \emph{Advances in Neural
  Information Processing Systems 34: Annual Conference on Neural Information
  Processing Systems 2021, NeurIPS 2021, December 6-14, 2021, virtual}, pages
  17981--17993, 2021.

\bibitem[Azqadan et~al.(2023)Azqadan, Jahed, and Arami]{azqadan2023predictive}
Erfan Azqadan, Hamid Jahed, and Arash Arami.
\newblock Predictive microstructure image generation using denoising diffusion
  probabilistic models.
\newblock \emph{Acta Materialia}, 261:\penalty0 119406, 2023.
\newblock ISSN 1359-6454.
\newblock \doi{https://doi.org/10.1016/j.actamat.2023.119406}.
\newblock URL
  \url{https://www.sciencedirect.com/science/article/pii/S135964542300736X}.

\bibitem[Bansal et~al.(2022)Bansal, Borgnia, Chu, Li, Kazemi, Huang, Goldblum,
  Geiping, and Goldstein]{Bansal2022ColdNoise}
Arpit Bansal, Eitan Borgnia, Hong-Min Chu, Jie~S. Li, Hamid Kazemi, Furong
  Huang, Micah Goldblum, Jonas Geiping, and Tom Goldstein.
\newblock Cold diffusion: Inverting arbitrary image transforms without noise,
  2022.

\bibitem[Bear(2013)]{JacobBear}
Jacob Bear.
\newblock \emph{Dynamics of fluids in porous media}.
\newblock Courier Corporation, 2013.

\bibitem[Blunt et~al.(2013)Blunt, Bijeljic, Dong, Gharbi, Iglauer, Mostaghimi,
  Paluszny, and Pentland]{blunt2013}
Martin~J Blunt, Branko Bijeljic, Hu~Dong, Oussama Gharbi, Stefan Iglauer,
  Peyman Mostaghimi, Adriana Paluszny, and Christopher Pentland.
\newblock Pore-scale imaging and modelling.
\newblock \emph{Advances in Water resources}, 51:\penalty0 197--216, 2013.

\bibitem[Boone(2014)]{carbonate_m}
MA~Boone.
\newblock {3D} mapping of water in oolithic limestone at atmospheric and vacuum
  saturation using x-ray micro-ct differential imaging.
\newblock \emph{Materials Characterization}, 97:\penalty0 150--160, 2014.

\bibitem[Bultreys et~al.(2016)Bultreys, Stappen, Kock, Boever, Boone,
  Hoorebeke, and Cnudde]{carbonate_s}
Tom Bultreys, Jeroen~Van Stappen, Tim~De Kock, Wesley~De Boever, Marijn~A
  Boone, Luc~Van Hoorebeke, and Veerle Cnudde.
\newblock Investigating the relative permeability behavior of
  microporosity-rich carbonates and tight sandstones with multiscale pore
  network models.
\newblock \emph{Journal of Geophysical Research: Solid Earth}, 121\penalty0
  (11):\penalty0 7929--7945, 2016.

\bibitem[Calcagnotto et~al.(2011)Calcagnotto, Adachi, Ponge, and
  Raabe]{calcagnotto2011deformation}
Marion Calcagnotto, Yoshitaka Adachi, Dirk Ponge, and Dierk Raabe.
\newblock Deformation and fracture mechanisms in fine-and ultrafine-grained
  ferrite/martensite dual-phase steels and the effect of aging.
\newblock \emph{Acta Materialia}, 59\penalty0 (2):\penalty0 658--670, 2011.

\bibitem[Campbell et~al.(2022)Campbell, Benton, De~Bortoli, Rainforth,
  Deligiannidis, and Doucet]{campbell2022}
Andrew Campbell, Joe Benton, Valentin De~Bortoli, Thomas Rainforth, George
  Deligiannidis, and Arnaud Doucet.
\newblock A continuous time framework for discrete denoising models.
\newblock In S.~Koyejo, S.~Mohamed, A.~Agarwal, D.~Belgrave, K.~Cho, and A.~Oh,
  editors, \emph{Advances in Neural Information Processing Systems}, volume~35,
  pages 28266--28279. Curran Associates, Inc., 2022.

\bibitem[Cao et~al.(2005)Cao, Gillespie, and
  Petzold]{caoAvoidingNegativePopulations2005}
Yang Cao, Daniel~T. Gillespie, and Linda~R. Petzold.
\newblock Avoiding negative populations in explicit {{Poisson}} tau-leaping.
\newblock \emph{The Journal of Chemical Physics}, 123\penalty0 (5):\penalty0
  054104, August 2005.
\newblock ISSN 0021-9606.
\newblock \doi{10.1063/1.1992473}.

\bibitem[Cao et~al.(2006)Cao, Gillespie, and Petzold]{caoEfficientStepSize2006}
Yang Cao, Daniel~T. Gillespie, and Linda~R. Petzold.
\newblock Efficient step size selection for the tau-leaping simulation method.
\newblock \emph{The Journal of Chemical Physics}, 124\penalty0 (4):\penalty0
  044109, January 2006.
\newblock ISSN 0021-9606.
\newblock \doi{10.1063/1.2159468}.

\bibitem[Chatterjee et~al.(2005)Chatterjee, Vlachos, and
  Katsoulakis]{chatterjeeBinomialDistributionBased2005}
Abhijit Chatterjee, Dionisios~G. Vlachos, and Markos~A. Katsoulakis.
\newblock Binomial distribution based {$\tau$}-leap accelerated stochastic
  simulation.
\newblock \emph{The Journal of Chemical Physics}, 122\penalty0 (2), January
  2005.
\newblock ISSN 0021-9606.
\newblock \doi{10.1063/1.1833357}.

\bibitem[Chung and Ye(2022)]{chungScorebasedDiffusionModels2022}
Hyungjin Chung and Jong~Chul Ye.
\newblock Score-based diffusion models for accelerated {{MRI}}.
\newblock \emph{Medical Image Analysis}, 80:\penalty0 102479, August 2022.
\newblock ISSN 1361-8415.
\newblock \doi{10.1016/j.media.2022.102479}.

\bibitem[Chung et~al.(2022{\natexlab{a}})Chung, Kim, Mc{C}ann, Klasky, and
  Ye]{chungDiffusionPosteriorSampling2022}
Hyungjin Chung, Jeongsol Kim, Michael~Thompson Mc{C}ann, Marc~Louis Klasky, and
  Jong~Chul Ye.
\newblock Diffusion {{Posterior Sampling}} for {{General Noisy Inverse
  Problems}}.
\newblock In \emph{The {{Eleventh International Conference}} on {{Learning
  Representations}}}, September 2022{\natexlab{a}}.

\bibitem[Chung et~al.(2022{\natexlab{b}})Chung, Sim, Ryu, and
  Ye]{chungImprovingDiffusionModels2022}
Hyungjin Chung, Byeongsu Sim, Dohoon Ryu, and Jong~Chul Ye.
\newblock Improving {{Diffusion Models}} for {{Inverse Problems}} using
  {{Manifold Constraints}}.
\newblock In \emph{Advances in {{Neural Information Processing Systems}}},
  October 2022{\natexlab{b}}.

\bibitem[Chung et~al.(2022{\natexlab{c}})Chung, Sim, and
  Ye]{chungComeCloserDiffuseFasterAcceleratingConditional2022}
Hyungjin Chung, Byeongsu Sim, and Jong~Chul Ye.
\newblock Come-{{Closer-Diffuse-Faster}}: {{Accelerating Conditional Diffusion
  Models}} for {{Inverse Problems}} through {{Stochastic Contraction}}.
\newblock In \emph{2022 {{IEEE}}/{{CVF Conference}} on {{Computer Vision}} and
  {{Pattern Recognition}} ({{CVPR}})}, pages 12403--12412, June
  2022{\natexlab{c}}.
\newblock \doi{10.1109/CVPR52688.2022.01209}.

\bibitem[Cnudde and Boone(2013)]{cnudde2013high}
Veerle Cnudde and Matthieu~Nicolaas Boone.
\newblock High-resolution x-ray computed tomography in geosciences: A review of
  the current technology and applications.
\newblock \emph{Earth-Science Reviews}, 123:\penalty0 1--17, 2013.

\bibitem[Corbella et~al.(2022)Corbella, Spencer, and
  Roberts]{corbellaAutomaticZigZagSampling2022a}
Alice Corbella, Simon E.~F. Spencer, and Gareth~O. Roberts.
\newblock Automatic {{Zig-Zag}} sampling in practice.
\newblock \emph{Statistics and Computing}, 32\penalty0 (6):\penalty0 107,
  November 2022.
\newblock ISSN 1573-1375.
\newblock \doi{10.1007/s11222-022-10142-x}.

\bibitem[Courant et~al.(1928)Courant, Friedrichs, and
  Lewy]{courantUberPartiellenDifferenzengleichungen1928}
R.~Courant, K.~Friedrichs, and H.~Lewy.
\newblock {\"U}ber die partiellen {{Differenzengleichungen}} der mathematischen
  {{Physik}}.
\newblock \emph{Mathematische Annalen}, 100\penalty0 (1):\penalty0 32--74,
  December 1928.
\newblock ISSN 1432-1807.
\newblock \doi{10.1007/BF01448839}.

\bibitem[Deng et~al.(2021)Deng, Lin, Huang, Meng, Zhong, Ma, Zhou, Shen, Ding,
  and Huang]{deng2021recent}
Zhe Deng, Xing Lin, Zhenyu Huang, Jintao Meng, Yun Zhong, Guangting Ma,
  Yu~Zhou, Yue Shen, Han Ding, and Yunhui Huang.
\newblock Recent progress on advanced imaging techniques for lithium-ion
  batteries.
\newblock \emph{Advanced Energy Materials}, 11\penalty0 (2):\penalty0 2000806,
  2021.

\bibitem[Du et~al.(2024)Du, Parikh, Fan, Liu, and
  Wang]{duCoNFiLDConditionalNeural2024}
Pan Du, Meet~Hemant Parikh, Xiantao Fan, Xin-Yang Liu, and Jian-Xun Wang.
\newblock {{CoNFiLD}}: {{Conditional Neural Field Latent Diffusion Model
  Generating Spatiotemporal Turbulence}}, March 2024.

\bibitem[Duquesnoy et~al.(2023)Duquesnoy, Liu, Dominguez, Kumar, Ayerbe, and
  Franco]{duquesnoy2023machine}
Marc Duquesnoy, Chaoyue Liu, Diana~Zapata Dominguez, Vishank Kumar, Elixabete
  Ayerbe, and Alejandro~A Franco.
\newblock Machine learning-assisted multi-objective optimization of battery
  manufacturing from synthetic data generated by physics-based simulations.
\newblock \emph{Energy Storage Materials}, 56:\penalty0 50--61, 2023.

\bibitem[Düreth et~al.(2023)Düreth, Seibert, Rücker, Handford, Kästner, and
  Gude]{dureth2023conditional}
Christian Düreth, Paul Seibert, Dennis Rücker, Stephanie Handford, Markus
  Kästner, and Maik Gude.
\newblock Conditional diffusion-based microstructure reconstruction.
\newblock \emph{Materials Today Communications}, 35:\penalty0 105608, 2023.
\newblock ISSN 2352-4928.
\newblock \doi{https://doi.org/10.1016/j.mtcomm.2023.105608}.
\newblock URL
  \url{https://www.sciencedirect.com/science/article/pii/S2352492823002982}.

\bibitem[Einstein(1905)]{einsteinUeberMolekularkinetischenTheorie1905}
Albert Einstein.
\newblock {{\"U}ber die von der molekularkinetischen Theorie der W{\"a}rme
  geforderte Bewegung von in ruhenden Fl{\"u}ssigkeiten suspendierten
  Teilchen}.
\newblock \emph{Annalen der Physik}, vol. 4, t. 17, 1905.

\bibitem[Finzi et~al.(2023)Finzi, Boral, Wilson, Sha, and
  {Zepeda-Nunez}]{finziUserdefinedEventSampling2023a}
Marc~Anton Finzi, Anudhyan Boral, Andrew~Gordon Wilson, Fei Sha, and Leonardo
  {Zepeda-Nunez}.
\newblock User-defined {{Event Sampling}} and {{Uncertainty Quantification}} in
  {{Diffusion Models}} for {{Physical Dynamical Systems}}.
\newblock In \emph{Proceedings of the 40th {{International Conference}} on
  {{Machine Learning}}}, pages 10136--10152. PMLR, July 2023.

\bibitem[Gardiner(2009)]{gardinerStochasticMethodsHandbook2009}
Crispin~W. Gardiner.
\newblock \emph{Stochastic Methods: A Handbook for the Natural and Social
  Sciences}.
\newblock Number~13 in Springer Series in Synergetics. {Springer}, {Berlin
  Heidelberg}, 4th ed edition, 2009.
\newblock ISBN 978-3-642-08962-6 978-3-540-70712-7.

\bibitem[Gayon-Lombardo et~al.(2020)Gayon-Lombardo, Mosser, Brandon, and
  Cooper]{gayon2020pores}
Andrea Gayon-Lombardo, Lukas Mosser, Nigel~P Brandon, and Samuel~J Cooper.
\newblock Pores for thought: generative adversarial networks for stochastic
  reconstruction of 3d multi-phase electrode microstructures with periodic
  boundaries.
\newblock \emph{npj Computational Materials}, 6\penalty0 (1):\penalty0 82,
  2020.

\bibitem[Gillespie(1976)]{gillespieGeneralMethodNumerically1976a}
Daniel~T Gillespie.
\newblock A general method for numerically simulating the stochastic time
  evolution of coupled chemical reactions.
\newblock \emph{Journal of Computational Physics}, 22\penalty0 (4):\penalty0
  403--434, December 1976.
\newblock ISSN 0021-9991.
\newblock \doi{10.1016/0021-9991(76)90041-3}.

\bibitem[Gillespie(2001)]{tauLeaping21}
Daniel~T. Gillespie.
\newblock Approximate accelerated stochastic simulation of chemically reacting
  systems.
\newblock \emph{The Journal of Chemical Physics}, 115\penalty0 (4):\penalty0
  1716--1733, 2001.
\newblock \doi{https://doi.org/10.1063/1.1378322}.

\bibitem[Gillespie and Petzold(2003)]{gillespieImprovedLeapsizeSelection2003}
Daniel~T. Gillespie and Linda~R. Petzold.
\newblock Improved leap-size selection for accelerated stochastic simulation.
\newblock \emph{The Journal of Chemical Physics}, 119\penalty0 (16):\penalty0
  8229--8234, October 2003.
\newblock ISSN 0021-9606.
\newblock \doi{10.1063/1.1613254}.

\bibitem[Giuggioli(2020)]{giuggioliExactSpatiotemporalDynamics2020}
Luca Giuggioli.
\newblock Exact {{Spatiotemporal Dynamics}} of {{Confined Lattice Random
  Walks}} in {{Arbitrary Dimensions}}: {{A Century}} after {{Smoluchowski}} and
  {{P}}{\textbackslash}'olya.
\newblock \emph{Physical Review X}, 10\penalty0 (2):\penalty0 021045, May 2020.
\newblock \doi{10.1103/PhysRevX.10.021045}.

\bibitem[Gostick et~al.(2019)Gostick, Khan, Tranter, Kok, Agnaou, Sadeghi, and
  Jervis]{porespy}
Jeff~T Gostick, Zohaib~A Khan, Thomas~G Tranter, Matthew~DR Kok, Mehrez Agnaou,
  Mohammadamin Sadeghi, and Rhodri Jervis.
\newblock Porespy: A python toolkit for quantitative analysis of porous media
  images.
\newblock \emph{Journal of Open Source Software}, 4\penalty0 (37):\penalty0
  1296, 2019.

\bibitem[Gu et~al.(2018)Gu, Chen, Richmond, and Buehler]{micromaterials}
Grace~X Gu, Chun-Teh Chen, Deon~J Richmond, and Markus~J Buehler.
\newblock Bioinspired hierarchical composite design using machine learning:
  simulation, additive manufacturing, and experiment.
\newblock \emph{Materials horizons.}, 5\penalty0 (5):\penalty0 939--945, 2018.
\newblock ISSN 2051-6355 (electronic).

\bibitem[Ho et~al.(2020)Ho, Jain, and
  Abbeel]{hoDenoisingDiffusionProbabilistic2020}
Jonathan Ho, Ajay Jain, and Pieter Abbeel.
\newblock Denoising diffusion probabilistic models.
\newblock In H.~Larochelle, M.~Ranzato, R.~Hadsell, M.F. Balcan, and H.~Lin,
  editors, \emph{Advances in Neural Information Processing Systems}, volume~33,
  pages 6840--6851. Curran Associates, Inc., 2020.

\bibitem[Hoffman et~al.(2025)Hoffman, Diniz, Liu, Rodgers, Tran, and
  Fuge]{hoffman2025grain}
Nathan Hoffman, Cashen Diniz, Dehao Liu, Theron Rodgers, Anh Tran, and Mark
  Fuge.
\newblock Grainpaint: A multi-scale diffusion-based generative model for
  microstructure reconstruction of large-scale objects.
\newblock \emph{Acta Materialia}, 288:\penalty0 120784, 2025.
\newblock ISSN 1359-6454.
\newblock \doi{https://doi.org/10.1016/j.actamat.2025.120784}.
\newblock URL
  \url{https://www.sciencedirect.com/science/article/pii/S135964542500076X}.

\bibitem[Hoogeboom and Salimans(2022)]{hoogeboomBlurringDiffusionModels2022}
Emiel Hoogeboom and Tim Salimans.
\newblock Blurring {{Diffusion Models}}.
\newblock In \emph{The {{Eleventh International Conference}} on {{Learning
  Representations}}}, September 2022.

\bibitem[Hoogeboom et~al.(2021)Hoogeboom, Nielsen, Jaini, Forr\'{e}, and
  Welling]{hoogeboom2021argmax}
Emiel Hoogeboom, Didrik Nielsen, Priyank Jaini, Patrick Forr\'{e}, and Max
  Welling.
\newblock Argmax flows and multinomial diffusion: Learning categorical
  distributions.
\newblock In M.~Ranzato, A.~Beygelzimer, Y.~Dauphin, P.S. Liang, and J.~Wortman
  Vaughan, editors, \emph{Advances in Neural Information Processing Systems},
  volume~34, pages 12454--12465. Curran Associates, Inc., 2021.

\bibitem[Janner et~al.(2022)Janner, Du, Tenenbaum, and
  Levine]{janner2022planningdiffusionflexiblebehavior}
Michael Janner, Yilun Du, Joshua~B. Tenenbaum, and Sergey Levine.
\newblock Planning with diffusion for flexible behavior synthesis, 2022.
\newblock URL \url{https://arxiv.org/abs/2205.09991}.

\bibitem[Kench et~al.(2023)Kench, Squires, and Cooper]{kench2023taufactor}
Steve Kench, Isaac Squires, and Samuel Cooper.
\newblock Taufactor 2: A gpu accelerated python tool for microstructural
  analysis.
\newblock \emph{Journal of Open Source Software}, 8\penalty0 (88):\penalty0
  5358, 2023.

\bibitem[Kingma and Welling(2014)]{kingmaAutoencodingVariationalBayes2014}
Diederik~P. Kingma and Max Welling.
\newblock {Auto-Encoding Variational Bayes}.
\newblock In \emph{2nd International Conference on Learning Representations,
  {ICLR} 2014, Banff, AB, Canada, April 14-16, 2014, Conference Track
  Proceedings}, 2014.

\bibitem[Krizhevsky et~al.(2009)Krizhevsky, Hinton,
  et~al.]{krizhevsky2009learning}
Alex Krizhevsky, Geoffrey Hinton, et~al.
\newblock Learning multiple layers of features from tiny images.
\newblock 2009.

\bibitem[Lawler(2010)]{lawlerRandomWalkHeat2010}
Gregory~F. Lawler.
\newblock \emph{Random Walk and the Heat Equation}.
\newblock Number volume 55 in Student Mathematical Library. American
  Mathematical Society, Providence, R.I, 2010.
\newblock ISBN 978-0-8218-4829-6.

\bibitem[LeCun et~al.(2010)LeCun, Cortes, and Burges]{mnist}
Yann LeCun, Corinna Cortes, and Christopher~J.C. Burges.
\newblock Mnist handwritten digit database, 2010.
\newblock URL \url{http://yann.lecun.com/exdb/mnist/}.

\bibitem[Lee and Yun(2024)]{lee2024microstructure}
Kang-Hyun Lee and Gun~Jin Yun.
\newblock Microstructure reconstruction using diffusion-based generative
  models.
\newblock \emph{Mechanics of Advanced Materials and Structures}, 31\penalty0
  (18):\penalty0 4443--4461, 2024.

\bibitem[Lee et~al.(2024)Lee, Lim, and Yun]{lee2024data}
Kang-Hyun Lee, Hyoung~Jun Lim, and Gun~Jin Yun.
\newblock A data-driven framework for designing microstructure of
  multifunctional composites with deep-learned diffusion-based generative
  models.
\newblock \emph{Engineering Applications of Artificial Intelligence},
  129:\penalty0 107590, 2024.
\newblock ISSN 0952-1976.
\newblock \doi{https://doi.org/10.1016/j.engappai.2023.107590}.
\newblock URL
  \url{https://www.sciencedirect.com/science/article/pii/S0952197623017748}.

\bibitem[Leonard(1948)]{simplifiedporositymeasurements}
RW~Leonard.
\newblock Simplified porosity measurements.
\newblock \emph{The Journal of the Acoustical Society of America}, 20\penalty0
  (1):\penalty0 39--41, 1948.

\bibitem[Lipman et~al.(2022)Lipman, Chen, {Ben-Hamu}, Nickel, and
  Le]{lipmanFlowMatchingGenerative2022}
Yaron Lipman, Ricky T.~Q. Chen, Heli {Ben-Hamu}, Maximilian Nickel, and Matthew
  Le.
\newblock Flow {{Matching}} for {{Generative Modeling}}.
\newblock In \emph{The {{Eleventh International Conference}} on {{Learning
  Representations}}}, September 2022.

\bibitem[Liu et~al.(2015)Liu, Luo, Wang, and Tang]{celebA}
Ziwei Liu, Ping Luo, Xiaogang Wang, and Xiaoou Tang.
\newblock Deep learning face attributes in the wild.
\newblock \emph{Proceedings of the IEEE International Conference on Computer
  Vision (ICCV)}, pages 3730--3738, 2015.

\bibitem[Lou et~al.(2024)Lou, Meng, and
  Ermon]{louDiscreteDiffusionLanguage2023}
Aaron Lou, Chenlin Meng, and Stefano Ermon.
\newblock Discrete diffusion modeling by estimating the ratios of the data
  distribution.
\newblock In Ruslan Salakhutdinov, Zico Kolter, Katherine Heller, Adrian
  Weller, Nuria Oliver, Jonathan Scarlett, and Felix Berkenkamp, editors,
  \emph{Proceedings of the 41st International Conference on Machine Learning},
  volume 235 of \emph{Proceedings of Machine Learning Research}, pages
  32819--32848. PMLR, 21--27 Jul 2024.
\newblock URL \url{https://proceedings.mlr.press/v235/lou24a.html}.

\bibitem[Lyu and Ren(2024)]{Lyu2024Microstructure}
Xianrui Lyu and Xiaodan Ren.
\newblock Microstructure reconstruction of 2d/3d random materials via
  diffusion-based deep generative models.
\newblock \emph{Scientific Reports}, 14\penalty0 (1):\penalty0 5041, 2024.

\bibitem[Montoya et~al.(2021)Montoya, Du, Gianforcaro, Orrego, Yang, and
  Lelkes]{Montoyabone2021}
Carolina Montoya, Yu~Du, Anthony~L. Gianforcaro, Santiago Orrego, Maobin Yang,
  and Peter~I. Lelkes.
\newblock On the road to smart biomaterials for bone research: definitions,
  concepts, advances, and outlook.
\newblock \emph{Bone Research}, 9\penalty0 (1):\penalty0 12, February 2021.
\newblock ISSN 2095-6231.
\newblock \doi{10.1038/s41413-020-00131-z}.
\newblock URL \url{https://doi.org/10.1038/s41413-020-00131-z}.

\bibitem[Neumann et~al.(2021)Neumann, Barsi-Andreeta, Lucas-Oliveira, Barbalho,
  Trevizan, Bonagamba, and Steiner]{berea}
Rodrigo~F Neumann, Mariane Barsi-Andreeta, Everton Lucas-Oliveira, Hugo
  Barbalho, Willian~A Trevizan, Tito~J Bonagamba, and Mathias~B Steiner.
\newblock High accuracy capillary network representation in digital rock
  reveals permeability scaling functions.
\newblock \emph{Scientific reports}, 11\penalty0 (1):\penalty0 11370, 2021.

\bibitem[Nichol and Dhariwal(2021)]{nicholImprovedDenoisingDiffusion2021}
Alexander~Quinn Nichol and Prafulla Dhariwal.
\newblock Improved denoising diffusion probabilistic models.
\newblock In Marina Meila and Tong Zhang, editors, \emph{Proceedings of the
  38th International Conference on Machine Learning}, volume 139 of
  \emph{Proceedings of Machine Learning Research}, pages 8162--8171. PMLR,
  18--24 Jul 2021.

\bibitem[{\O}ren and Bakke(2002)]{processbased}
P{\aa}l-Eric {\O}ren and Stig Bakke.
\newblock Process based reconstruction of sandstones and prediction of
  transport properties.
\newblock \emph{Transport in porous media}, 46\penalty0 (2):\penalty0 311--343,
  2002.

\bibitem[Passey et~al.(1990)Passey, Creaney, Kulla, Moretti, and
  Stroud]{practicalporosity}
QR~Passey, S~Creaney, JB~Kulla, FJ~Moretti, and JD~Stroud.
\newblock A practical model for organic richness from porosity and resistivity
  logs.
\newblock \emph{AAPG bulletin}, 74\penalty0 (12):\penalty0 1777--1794, 1990.

\bibitem[Phogat et~al.(2024)Phogat, Sharma, Jha, and
  Singh]{phogat2024microstructural}
Peeyush Phogat, Shreya Sharma, Ranjana Jha, and Sukhvir Singh.
\newblock Microstructural influence on electrochemical devices.
\newblock In \emph{Electrochemical Devices: Principles to Applications}, pages
  257--306. Springer, 2024.

\bibitem[Ren et~al.(2024)Ren, Chen, Rotskoff, and
  Ying]{renHowDiscreteContinuous2024}
Yinuo Ren, Haoxuan Chen, Grant~M. Rotskoff, and Lexing Ying.
\newblock How {{Discrete}} and {{Continuous Diffusion Meet}}: {{Comprehensive
  Analysis}} of {{Discrete Diffusion Models}} via a {{Stochastic Integral
  Framework}}, October 2024.

\bibitem[Risken(1984)]{riskenFokkerPlanckEquation1984}
Hermann Risken.
\newblock \emph{The {{Fokker-Planck Equation}}}.
\newblock Springer-Verlag Berlin Heidelberg, 1984.
\newblock ISBN 978-3-642-96809-9.

\bibitem[Rissanen et~al.(2023)Rissanen, Heinonen, and Solin]{heat_eq}
Severi Rissanen, Markus Heinonen, and Arno Solin.
\newblock Generative modelling with inverse heat dissipation.
\newblock In \emph{The Eleventh International Conference on Learning
  Representations}, 2023.
\newblock \doi{10.48550/arxiv.2206.13397}.

\bibitem[Santos and Lin(2023)]{santosUnderstandingDenoisingDiffusion2023}
Javier~E. Santos and Yen~Ting Lin.
\newblock Understanding {{Denoising Diffusion Probabilistic Models}} and their
  {{Noise Schedules}} via the {{Ornstein--Uhlenbeck Process}}.
\newblock In \emph{{{NeurIPS}} 2023 {{Workshop}} on {{Diffusion Models}}},
  October 2023.

\bibitem[Santos et~al.(2023)Santos, Fox, Lubbers, and Lin]{blackout}
Javier~E. Santos, Zachary~R. Fox, Nicholas Lubbers, and Yen~Ting Lin.
\newblock Blackout diffusion: Generative diffusion models in discrete-state
  spaces.
\newblock In Andreas Krause, Emma Brunskill, Kyunghyun Cho, Barbara Engelhardt,
  Sivan Sabato, and Jonathan Scarlett, editors, \emph{Proceedings of the 40th
  International Conference on Machine Learning}, volume 202 of
  \emph{Proceedings of Machine Learning Research}, pages 9034--9059. PMLR,
  23--29 Jul 2023.
\newblock URL \url{https://proceedings.mlr.press/v202/santos23a.html}.

\bibitem[Shi et~al.(2024)Shi, Han, Wang, Doucet, and
  Titsias]{shi2024simplified}
Jiaxin Shi, Kehang Han, Zhe Wang, Arnaud Doucet, and Michalis Titsias.
\newblock Simplified and generalized masked diffusion for discrete data.
\newblock In A.~Globerson, L.~Mackey, D.~Belgrave, A.~Fan, U.~Paquet,
  J.~Tomczak, and C.~Zhang, editors, \emph{Advances in Neural Information
  Processing Systems}, volume~37, pages 103131--103167. Curran Associates,
  Inc., 2024.
\newblock URL
  \url{https://proceedings.neurips.cc/paper_files/paper/2024/file/bad233b9849f019aead5e5cc60cef70f-Paper-Conference.pdf}.

\bibitem[Simon and Gogotsi(2008)]{Simonenergystorage}
Patrice Simon and Yury Gogotsi.
\newblock Materials for electrochemical capacitors.
\newblock \emph{Nature Materials}, 7\penalty0 (11):\penalty0 845--854, 2008.
\newblock ISSN 1476-4660.
\newblock \doi{10.1038/nmat2297}.
\newblock URL \url{https://doi.org/10.1038/nmat2297}.

\bibitem[Sohl-Dickstein et~al.(2015)Sohl-Dickstein, Weiss, Maheswaranathan, and
  Ganguli]{sohl-dicksteinDeepUnsupervisedLearning2015}
Jascha Sohl-Dickstein, Eric Weiss, Niru Maheswaranathan, and Surya Ganguli.
\newblock Deep unsupervised learning using nonequilibrium thermodynamics.
\newblock In Francis Bach and David Blei, editors, \emph{Proceedings of the
  32nd International Conference on Machine Learning}, volume~37 of
  \emph{Proceedings of Machine Learning Research}, pages 2256--2265, Lille,
  France, 07--09 Jul 2015. PMLR.
\newblock URL \url{https://proceedings.mlr.press/v37/sohl-dickstein15.html}.

\bibitem[Song et~al.(2021{\natexlab{a}})Song, Meng, and
  Ermon]{song2020denoising}
Jiaming Song, Chenlin Meng, and Stefano Ermon.
\newblock Denoising diffusion implicit models.
\newblock In \emph{International Conference on Learning Representations},
  2021{\natexlab{a}}.

\bibitem[Song and Ermon(2019)]{songGenerativeModelingEstimating2020}
Yang Song and Stefano Ermon.
\newblock Generative modeling by estimating gradients of the data distribution.
\newblock In \emph{Advances in Neural Information Processing Systems}, pages
  11895--11907, 2019.

\bibitem[Song et~al.(2021{\natexlab{b}})Song, {Sohl-Dickstein}, Kingma, Kumar,
  Ermon, and Poole]{songScoreBasedGenerativeModeling2021a}
Yang Song, Jascha {Sohl-Dickstein}, Diederik~P. Kingma, Abhishek Kumar, Stefano
  Ermon, and Ben Poole.
\newblock Score-{{Based Generative Modeling}} through {{Stochastic Differential
  Equations}}, February 2021{\natexlab{b}}.

\bibitem[Sun et~al.(2022)Sun, Yu, Dai, Schuurmans, and
  Dai]{sunScorebasedContinuoustimeDiscrete2022}
Haoran Sun, Lijun Yu, Bo~Dai, Dale Schuurmans, and Hanjun Dai.
\newblock Score-based {{Continuous-time Discrete Diffusion Models}}.
\newblock In \emph{The {{Eleventh International Conference}} on {{Learning
  Representations}}}, September 2022.

\bibitem[Tian and Burrage(2004)]{tianBinomialLeapMethods2004}
Tianhai Tian and Kevin Burrage.
\newblock Binomial leap methods for simulating stochastic chemical kinetics.
\newblock \emph{The Journal of Chemical Physics}, 121\penalty0 (21):\penalty0
  10356--10364, December 2004.
\newblock ISSN 0021-9606.
\newblock \doi{10.1063/1.1810475}.

\bibitem[Torquato(2002)]{torquato2002random}
Salvatore Torquato.
\newblock \emph{Random heterogeneous materials: microstructure and macroscopic
  properties}.
\newblock Springer, 2002.

\bibitem[Uhlenbeck and Ornstein(1930)]{OU}
G.~E. Uhlenbeck and L.~S. Ornstein.
\newblock On the theory of the brownian motion.
\newblock \emph{Phys. Rev.}, 36:\penalty0 823--841, Sep 1930.
\newblock \doi{10.1103/PhysRev.36.823}.
\newblock URL \url{https://link.aps.org/doi/10.1103/PhysRev.36.823}.

\bibitem[Usseglio-Viretta et~al.(2018)Usseglio-Viretta, Colclasure, Mistry,
  Claver, Pouraghajan, Finegan, Heenan, Abraham, Mukherjee, Wheeler,
  et~al.]{usseglio2018resolving}
Francois~LE Usseglio-Viretta, Andrew Colclasure, Aashutosh~N Mistry, Koffi
  Pierre~Yao Claver, Fezzeh Pouraghajan, Donal~P Finegan, Thomas~MM Heenan,
  Daniel Abraham, Partha~P Mukherjee, Dean Wheeler, et~al.
\newblock Resolving the discrepancy in tortuosity factor estimation for li-ion
  battery electrodes through micro-macro modeling and experiment.
\newblock \emph{Journal of The Electrochemical Society}, 165\penalty0
  (14):\penalty0 A3403--A3426, 2018.

\bibitem[{Van Kampen}(2007)]{vanKampen}
N~G {Van Kampen}.
\newblock \emph{Stochastic {{Processes}} in {{Physics}} and {{Chemistry}}}.
\newblock {Elsevier Science B.V.}, {Amsterdam}, 2007.

\bibitem[Wang et~al.(2004)Wang, Bovik, Sheikh, and Simoncelli]{SSIM}
Zhou Wang, A.C. Bovik, H.R. Sheikh, and E.P. Simoncelli.
\newblock Image quality assessment: from error visibility to structural
  similarity.
\newblock \emph{IEEE Transactions on Image Processing}, 13\penalty0
  (4):\penalty0 600--612, 2004.
\newblock \doi{10.1109/TIP.2003.819861}.

\bibitem[Weber and Frey(2017)]{weberMasterEquationsTheory2017}
Markus~F. Weber and Erwin Frey.
\newblock Master equations and the theory of stochastic path integrals.
\newblock \emph{Reports on Progress in Physics}, 80\penalty0 (4), 2017.
\newblock ISSN 00344885.
\newblock \doi{10.1088/1361-6633/aa5ae2}.

\bibitem[Winkler et~al.(2024)Winkler, Richter, and
  Opper]{bridgingDiscreteContinuous24}
Ludwig Winkler, Lorenz Richter, and Manfred Opper.
\newblock Bridging discrete and continuous state spaces: exploring the
  ehrenfest process in time-continuous diffusion models.
\newblock In \emph{Proceedings of the 41st International Conference on Machine
  Learning}, ICML'24. JMLR.org, 2024.

\bibitem[Xiong et~al.(2010)Xiong, Selleby, Chen, Odqvist, and
  Du]{xiong2010phase}
Wei Xiong, Malin Selleby, Qing Chen, Joakim Odqvist, and Yong Du.
\newblock Phase equilibria and thermodynamic properties in the fe-cr system.
\newblock \emph{Critical Reviews in Solid State and Materials Sciences},
  35\penalty0 (2):\penalty0 125--152, 2010.

\bibitem[Zhu et~al.(2025)Zhu, Bijeljic, and Blunt]{bluntdiffusion}
L.~Zhu, B.~Bijeljic, and M.J. Blunt.
\newblock Diffusion model-based generation of three-dimensional multiphase
  pore-scale images.
\newblock \emph{Transport in Porous Media}, 152:\penalty0 22, 2025.
\newblock \doi{10.1007/s11242-025-02158-4}.
\newblock URL \url{https://doi.org/10.1007/s11242-025-02158-4}.

\end{thebibliography}

\newpage
\appendix

\section{Deriving reverse-time transition rates} \label{app:reverseRate}
Here, we derive the reverse-time transition rate. Because the particles are moving independently, it is sufficient to discuss $n$ particles colocalized at $(x,y)$ in channel $c$, and the conclusion applies to other locations and color channels. For brevity, we will drop the $(x,y,c)$ dependence in this section when the context is clear. Let us index the particles by $i=1\ldots n = \left[I_t\right]_{x,y,c}$. For each of the $n$ particles, the reverse-time transition rates moving to $(x+{\bar{\nu}}_x, y+{\bar{\nu}}_y)$, where $\left({\bar{\nu}}_x, {\bar{\nu}}_y\right) \in  \left\{\left(-1,0\right), \left(1,0\right),\left(0,-1\right), \left(0,1\right)\right\}$ is
\begin{equation}
    \bar{r}^{[i]}_{{\bar{\nu}}}(t) = r\frac{p_t\left(x+{\bar{\nu}}_x, y+{\bar{\nu}}_y, c\vert x_0^{[i]}, y_0^{[i]}, c_0^{[i]}\right)}{p_t\left(x,y,c\vert x_0^{[i]}, y_0^{[i]}, c_0^{[i]}\right)}, \label{eq:reverseRate-app}
\end{equation}
according to the general theory of reverse-time dynamics for continuous-time Markov systems \cite{campbell2022,blackout}. We now perform the survival analysis for the inhomogeneous process. Within time ${\rm d} t$, the probability that particle $i$ leaves $(x,y,c)$ and moves to $(x+{\bar{\nu}}_x, y+{\bar{\nu}}_y, c)$ is $\bar{r}_{\bar{\nu}}^{i} \left(t\right) {\rm d}t + \mathcal{O}\left({\rm d} t^2\right)$. As such, the probability of the particle remains at $(x,y,c)$ at time $t - {\rm d}t $ is $1- \sum_{{\bar{\nu}}} \bar{r}_{\bar{\nu}}^{[i]} \left(t\right) {\rm d}t + \mathcal{O}\left({\rm d} t^2\right) $. Thanks to the independence between the particle dynamics, the probability of \textit{all} $n$ particles remaining at $(x,y,c)$ at time $t - {\rm d}t $ (recall that we are evolving the reverse-time dynamics) is $1-\sum_{i=1}^n \sum_{{\bar{\nu}}} \bar{r}_{\bar{\nu}}^{[i]} \left(t\right) {\rm d}t + \mathcal{O}\left({\rm d} t^2\right) $. Then, the probability of \textit{no} particle leaving at a previous time $t-\Delta t$, where $\Delta t:=N {\rm d}t $ is
\begin{equation}
    \prod_{k=1}^N \left[1-\sum_{i=1}^n \sum_{{\bar{\nu}}} \bar{r}_{\bar{\nu}}^{[i]} \left( t-\left(k-1\right) {\rm d} t\right) {\rm d}t\right] + \mathcal{O}\left({\rm d} t^2\right),
\end{equation}
which by sending ${\rm d} t \downarrow 0$ leads to the continuous-time survival function:
\begin{equation}
    \mathbb{P}\left\{ \mathcal{T} > t\right\}= \exp \left[ - \int_0^t \sum_{i,{\bar{\nu}}} \bar{r}^{[i]}_{\bar{\nu}}(t') {\rm d} t' \right],
\end{equation}
where $\mathcal{T}$ is the random time of the first particle moving out of $(x,y,c)$, the sum is over all possible directions and all particle index $i\in \left\{1\ldots n\right\}$. Identifying the total rate $\sum_{i,{\bar{\nu}}} \bar{r}^{[i]}_{\bar{\nu}}(t') {\rm d} t'$ and the reverse-time transition rate for each particle and in each direction, Eq.~\eqref{eq:reverseRate-app}, we arrived at Eq.~\eqref{eq:reverseRates}.  

\section{Training and generation algorithms.}

Algorithm~\ref{alg:training} gives the training algorithm using standard gradient descent techniques, and Algorithm~\ref{alg:inference} gives the inference algorithm used in this work.
\begin{algorithm}[tb]
   \caption{DSD training}
   \label{alg:training}
\begin{algorithmic}
    \STATE Given the full transition probabilities $p_t(x',y',c'\vert x, y, c)$
   \REPEAT
   \STATE $I_0 \leftarrow $ a sample drawn from the training set
   \STATE Draw an index $k$ from $\left\{1, \ldots T\right\}$ uniformly
   \FOR {Each discrete intensity unit in $[I_0]_{x,y,c}$}
        \STATE  Draw $(x',y',c') \sim p_t(x',y',c'\vert x,y,c)$
        \STATE Move the unit from $(x,y,c)$ to $(x',y',c')$
   \ENDFOR
   \STATE $I_{t_k} \leftarrow $ the corrupted image
   \STATE Compute the reverse transition rate Eq.~\eqref{eq:reverseRates}
   \IF{Using $L^1$ rate-matching}
   \STATE $\text{Loss} \leftarrow \sum_{x,y,c,\bar{\nu}} \left\vert \bar{r}^\text{NN} - \bar{r} \right\vert $ 
   \ELSIF{Using likelihood loss}
   \STATE $\text{Loss} \leftarrow -\log L$, defined in Eq.~\eqref{eq:likelihood} 
   \ENDIF
   \STATE Take a gradient step on $\nabla_\theta \text{Loss}$
   \UNTIL{Converged}
\end{algorithmic}
\end{algorithm}

\begin{algorithm}[tb]
   \caption{DSD inference}
   \label{alg:inference}
\begin{algorithmic}
   \STATE Given CFL condition number $\varepsilon <1$ and desired total intensities in the color channels, initiate an image $I_0$ with desired total intensities in the color channels
   \FOR {Each discrete intensity unit in $[I_0]_{x,y,c}$}
        \STATE  Draw $(x',y',c') \sim p_{1}(x',y',c'\vert x,y,c)$
        \STATE Move the unit from $(x,y,c)$ to $(x',y',c')$
   \ENDFOR
   \STATE $I_{1} \leftarrow $ the fully corrupted image, $t\leftarrow 1$
   \WHILE{$t>0$}
   \STATE Evaluate NN predicted reverse rates $\bar{r}^\text{NN}_{\bar{\nu},x,y,c}$
   \STATE $\tau \leftarrow \min \left\{t, \mathbf \varepsilon \min_{\bar{\nu}, x,y,c} \left(\bar{r}^\text{NN}_{\bar{\nu},x,y,c}\right)^{-1} \right\}$
   \FOR {each $(x,y,c)$}
   \STATE Sample total moving particles: 
   \STATE $n_\Sigma \sim \text{Binom}\left( \left[I_t\right]_{x,y,c}, \sum_{\bar{\nu}} \bar{r}^\text{NN}_{\bar{\nu},x,y,c} \right)$
   \STATE Sample a direction $\bar{\nu}$ for each moving particle:
   \STATE $n_{\bar{\nu}} \sim \text{Multinomial}\left(n_\Sigma, p_{\bar{\nu}} = \frac{\bar{r}^\text{NN}_{\bar{\nu},x,y,c} }{\sum_{\bar{\nu}'} \bar{r}^\text{NN}_{\bar{\nu}',x,y,c} }\right)$
   \STATE Move $n_{\bar{\nu}}$ intensity units to $(x+\bar{\nu}_x,y+\bar{\nu}_y)$
   \ENDFOR
   \STATE Advance time: $t \leftarrow t - \tau $
   \STATE $I_t \leftarrow $ the configuration after movements
   \ENDWHILE
\end{algorithmic}
\end{algorithm}

\onecolumn

\begin{figure*}[ht!]
    \centering
    \includegraphics[width=0.95\textwidth]{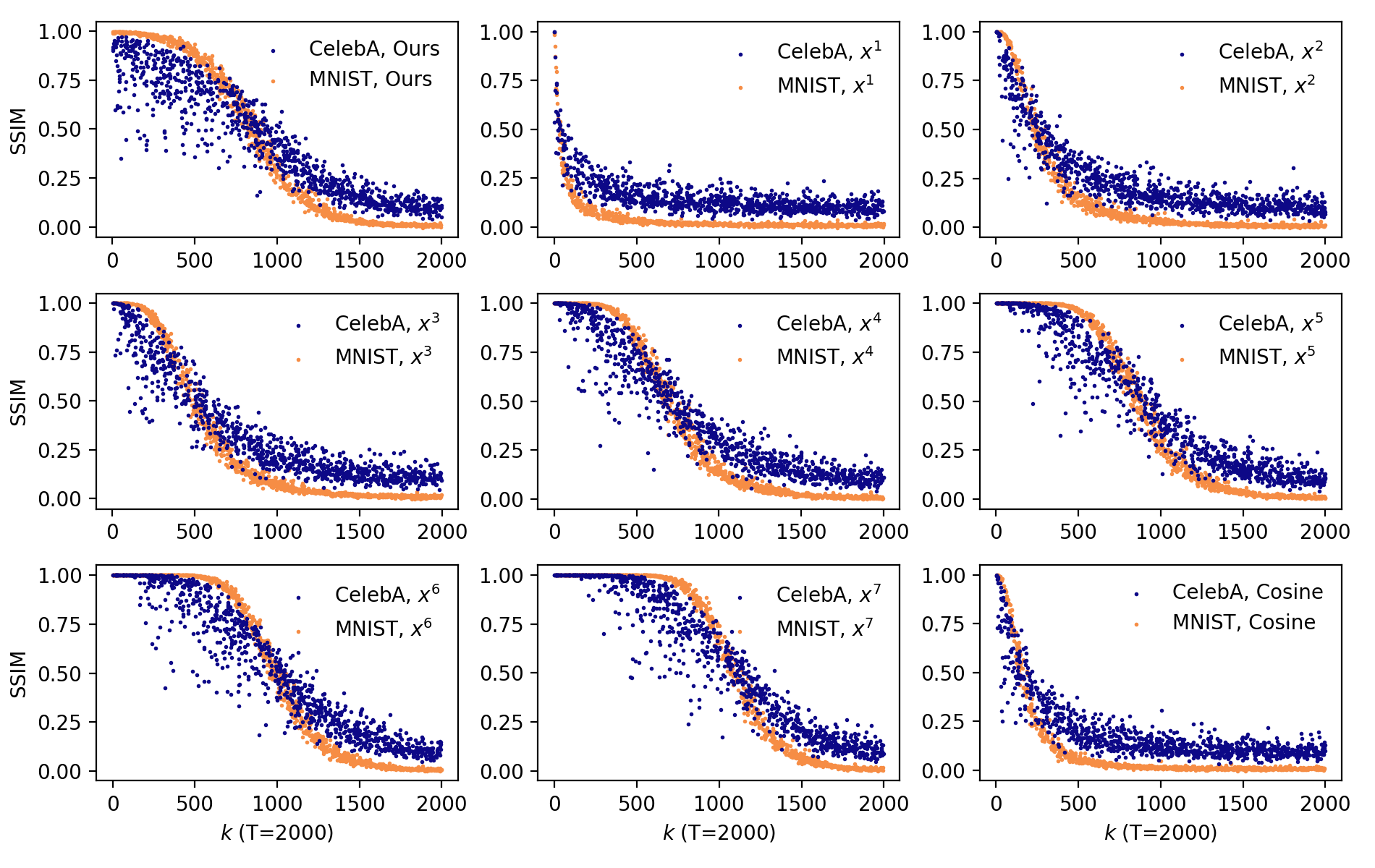}
    \caption{Structural Similarity Index Metric between the original and corrupted MNIST and CelebA images (1,000 samples evaluated at uniformly sampled random time $k\in \left\{1\ldots 2000\right\}$) with various noise scheduler. We fixed $r=120$ for MNIST and $r=200$ for CelebA in this analysis. For ours, we use Eq.~\eqref{eq:observationTimes} with $\tau_1=7.5$ and $\tau_2=2.5$. For the polynomials, $t_k= (k/T)^n$, $n=1\ldots 7$. For the cosine schedule, we use the heuristic formula given in \cite{nicholImprovedDenoisingDiffusion2021}. We remark that although the cosine schedule has been shown to be superior in previous studies \cite{nicholImprovedDenoisingDiffusion2021,santosUnderstandingDenoisingDiffusion2023}, the conclusion is based on the Ornstein--Uhlenbeck process, which is distinct from the spatial diffusion process \eqref{eq:forwardProcess}.}
    \label{fig:SSIM}
\end{figure*}

\section{Spectral representation of DSD with periodic boundary conditions}\label{app:DSD-FT}

\newcommand{\dd}{\text{d}}
Let us consider the two-dimensional DSD with periodic boundary conditions, whose master equation \cite{weberMasterEquationsTheory2017} is
\begin{equation}
\frac{\dd }{\dd t} p_{k,\ell}(t) = r p_{k-1,\ell}(t) + rp_{k+1,\ell}(t) + rp_{k,\ell-1}(t) + rp_{k,\ell+1}(t)  - 4 r p_{k,\ell}(t).
\end{equation}
Here, we use periodic boundary conditions, $p_{i\pm N,\ell} = p_{k,\ell}$, $p_{k,\ell\pm N} = p_{k,\ell}$ over the domain $(k,\ell)\in \left\{1, \ldots N_x \right\} \times  \left\{1, \ldots N_y \right\}$. 
We can formally treat $p_{k,l}$ as a rank-2 tensor. 
We can write the one-dimensional discrete-space Laplace operator:
\begin{equation}
    \left[\mathbf{L}_{x,y}\right]_{\mu,\nu} := \delta_{\mu+1,\nu} + \delta_{\mu-1,\nu} -2 \delta_{\mu,\nu}
\end{equation}
where $\mathbf{L}_x$ is $N_x\times N_x$ and $\mathbf{L}_y$ is $N_y \times N_y$, $\delta$ is the Kronecker delta. 
Then, we can write the two-dimensional master equation succinctly as
\begin{equation}
    \frac{\dd}{\dd t} \mathbf{p}(t) = r \left(\mathbf{L}_x \otimes I + I\otimes \mathbf{L}_y \right) \mathbf{p}(t).
\end{equation}

We aim to identify the spectral representation of the full operator $L$. To achieve this, we first define the two-dimensional discrete Fourier and inverse Fourier transformations:
\begin{subequations}\label{eq:plainFourierTransform}
    \begin{align}
    k_{m,n} (t) :={}& \frac{1}{\sqrt{N}} \sum_{k=1}^{N_x} \sum_{\ell=1}^{N_y} e^{-2\pi i \left(\frac{m k}{N_x} + \frac{n \ell}{N_y}\right) } p_{k,\ell}(t), \\
    p_{k,\ell} (t) :={}& \frac{1}{\sqrt{N}} \sum_{m=1}^{N_x} \sum_{n=1}^{N_y} e^{2\pi i\left(\frac{m k}{N_x} + \frac{n \ell}{N_y}\right) } k_{m,n}(t).
    \end{align}
\end{subequations}
To simplify the notation, we write $U_x$ and $U_y$ as the one-dimensional inverse discrete Fourier transformation 
\begin{equation}
    \left[\mathbf{U}_{x,y}\right]_{\mu,\nu}= \frac{1}{\sqrt{N}} e^{2 \pi i \frac{\mu \nu}{N_{x,y}} }.
\end{equation}
and again use the tensor notation. Then, Eqs.~\eqref{eq:plainFourierTransform} can be succinctly represented as
\begin{subequations}\label{eq:FourierTransform}
    \begin{align}
    \mathbf{k} (t) :={}& \left(\mathbf{U}_x \otimes I \right)^\dagger \left(I\otimes \mathbf{U}_y \right)^\dagger \mathbf{p} (t), \\
    \mathbf{p} (t) :={}& \left(\mathbf{U}_x \otimes I \right) \left(I\otimes \mathbf{U}_y \right) \mathbf{k} (t).
    \end{align}
\end{subequations}

The master equation in the spectral space can be derived now:
\begin{align}
    \frac{\dd}{\dd t} \mathbf{k} (t) :={}& \left(\mathbf{U}_x \otimes I \right)^\dagger \left(I\otimes \mathbf{U}_y \right)^\dagger \frac{\dd}{\dd t} \mathbf{p} (t) \nonumber \\
    = {}&r \left(\mathbf{U}_x \otimes I \right)^\dagger \left(I\otimes \mathbf{U}_y \right)^\dagger \left(\mathbf{L}_x \otimes I + I\otimes \mathbf{L}_y \right) \mathbf{p}(t) \nonumber \\
    = {}& r\left(\mathbf{U}_x \otimes I \right)^\dagger \left(I\otimes \mathbf{U}_y \right)^\dagger \left(\mathbf{L}_x \otimes I + I\otimes \mathbf{L}_y \right) \left(\mathbf{U}_x \otimes I \right) \left(I\otimes \mathbf{U}_y \right) \mathbf{k} (t) \nonumber \\
    ={}& r \left(\mathbf{U}_x^\dagger \mathbf{L}_x \mathbf{U}_x \otimes I + I\otimes \mathbf{U}_y^\dagger \mathbf{L}_y \mathbf{U}_y  \right) \mathbf{k} (t)
\end{align}
because $\mathbf{U}_x^\dagger \mathbf{U}_x=\mathbf{U}_y^\dagger \mathbf{U}_y =\mathbf{I}$. 

Let us now explicitly compute $\mathbf{U}_x^\dagger \mathbf{L}_x \mathbf{U}_x$:
\begin{align}
\left[\mathbf{U}_x^\dagger \mathbf{L}_x \mathbf{U}_x\right]_{k,\ell} ={}& \sum_{m=1}^{N_x} \sum_{n=1}^{N_x} \left[\mathbf{U}_x\right]_{k, m}^{\dagger} \left[\mathbf{L}_x\right]_{m,n} \left[\mathbf{U}_x\right]_{n, \ell} \nonumber \\
={}& \sum_{m=1}^{N_x} \sum_{n=1}^{N_x} \frac{1}{N} e^{2\pi i \frac{mk-n\ell}{N_x}} \left(\delta_{m+1,n}+\delta_{m-1,n} - 2 \delta_{m,n}\right)\nonumber \\
={}& \sum_{m=1}^{N_x} \frac{1}{N} \left(e^{2\pi i \frac{mk-\left(m+1\right)\ell}{N_x}} + e^{2\pi i \frac{mk-\left(m-1\right)\ell}{N_x}} - 2 e^{2\pi i \frac{mk-m\ell}{N_x}} \right) \nonumber \\
={}& \sum_{m=1}^{N_x} \frac{1}{N} e^{2\pi i \frac{m \left(k-\ell\right)}{N_x}} \left(e^{-\frac{2\pi i \ell}{N_x} } + e^{+\frac{2\pi i \ell}{N_x}} - 2 \right) \nonumber \\
={}& \left( 2 \cos \left( \frac{2 \pi \ell}{N_x} \right) - 2 \right) \delta_{k,\ell} =  4 \sin^2 \left( \frac{2 \pi \ell}{N_x} \right) \delta_{k,\ell} .
\end{align}
This shows that the process in the spectral space is diagonal with a cosine spectrum. The same analysis applies to $\mathbf{U}_y^\dagger \mathbf{L}_y \mathbf{U}_y$. Thus, element-wise, in the discrete Fourier space, we have
\begin{equation}
    \frac{\dd}{\dd t} k_{m,n} (t) = 4 r \left[ \sin^2 \left(\frac{\pi m}{N_x}\right)+ \sin^2 \left(\frac{\pi n}{N_y}\right)\right] k_{m,n}(t),
\end{equation}
which means
\begin{equation}
    k_{m,n} (t) = e^{4  \left[ \sin^2 \left(\frac{\pi m}{N_x}\right)+ \sin^2 \left(\frac{\pi n}{N_y}\right)\right] rt } k_{m,n} (0). 
\end{equation}

In practice, we can always shift the particle of interest at $t=0$ at the origin, i.e., $p_{0,0}(0)=1$ and $p_{i,j}=0$ if $(i,j)\ne (0,0)$. Then, $k_{m,n}(0) = \frac{1}{N}$ and consequently
\begin{equation}
    k_{m,n}(t) =  \frac{1}{N} e^{4  \left[ \sin^2 \left(\frac{\pi m}{N_x}\right)+ \sin^2 \left(\frac{\pi n}{N_y}\right)\right] rt },
\end{equation}
so the analytical solution of $p_{k, \ell}$ can be expressed as
\begin{equation}
    p_{k,\ell}(t) =  \frac{1}{\sqrt{N}^3}  \sum_{m=1}^{N_x} \sum_{n=1}^{N_y} e^{2\pi i\left(\frac{m k}{N_x} + \frac{n \ell}{N_y}\right) + 4  \left[ \sin^2 \left(\frac{\pi m}{N_x}\right)+ \sin^2 \left(\frac{\pi n}{N_y}\right)\right] rt}. 
\end{equation}
Note that the double sum can be carried out in one two-dimensional Fast Fourier Transformation. The above equation allows us to compute the transition probability efficiently.

\section{Human-centric datasets}
\label{sec:human-centric}

While our primary focus is on scientific data, we include results on standard vision benchmarks to contextualize DSD’s performance to the computer vision community.

\subsection{MNIST experiments}
\label{app:mnist}

\begin{figure*}[ht!]
    \centering
    \includegraphics[width=1\linewidth]{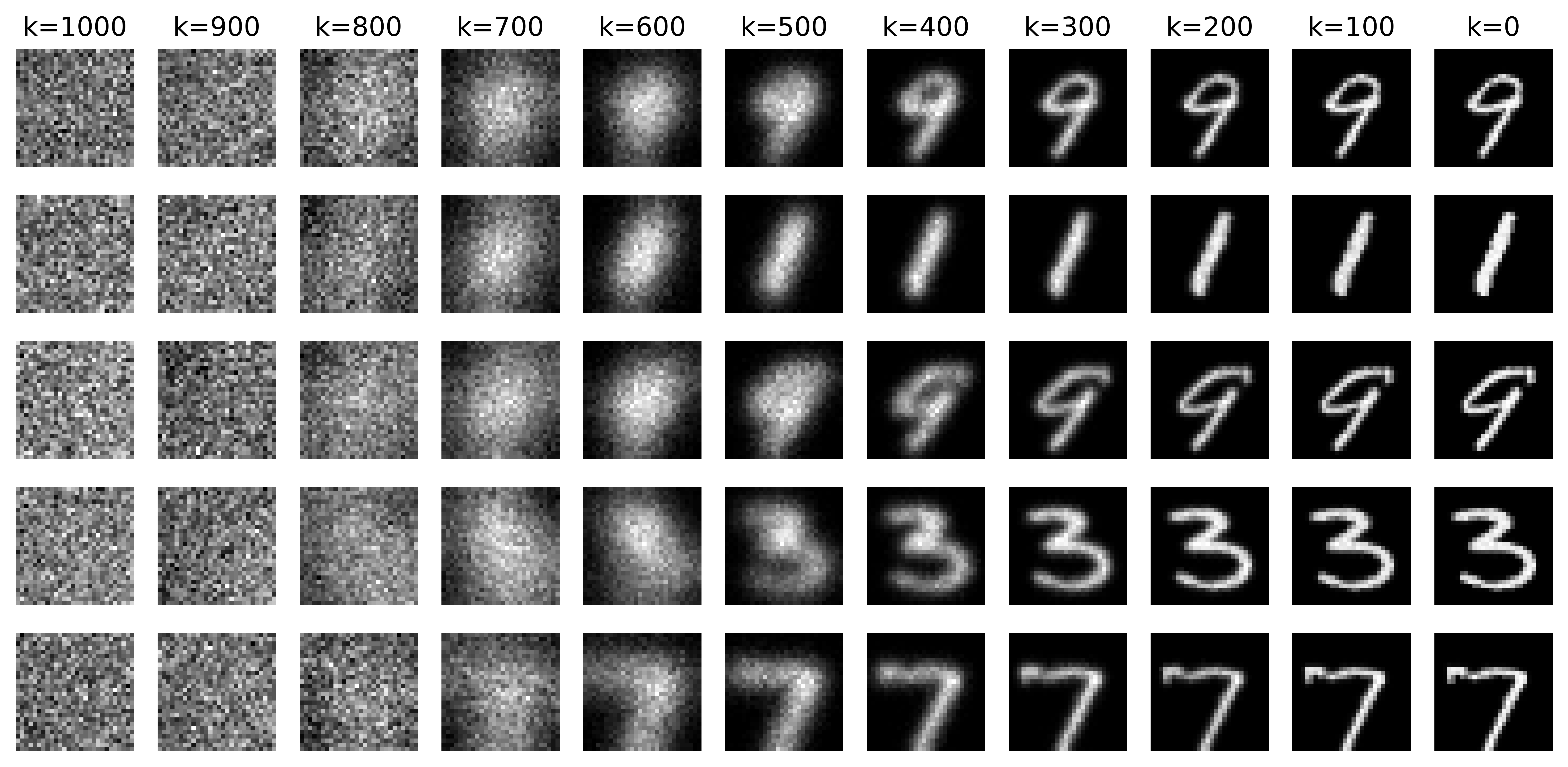}
    \caption{Unconditional MNIST generations.}
    \label{fig:mnist_unconditional}
\end{figure*}

As an initial validation, we trained an unconditional DSD model on MNIST and found that it reliably captures digit structure and diversity (Fig.~\ref{fig:mnist_unconditional}).

\begin{figure*}[ht!]
    \centering
    \includegraphics[width=.75\linewidth]{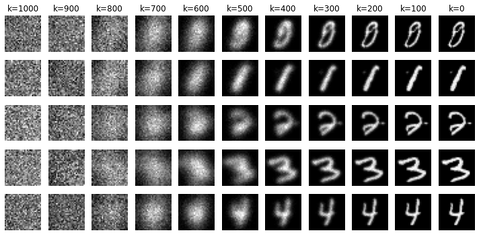}
    \caption{Class-conditional MNIST generations for digits 0 through 4. Each row corresponds to a specified target class.}
    \label{fig:mnist}
\end{figure*}

Then, in our additional MNIST experiments, we explored class-conditional and inpainting generation. These experiments are particularly notable due to their interactions with the intensity-preserving property of DSD. For class-conditioning, we introduced class embeddings into our model following the approach described in \cite{songScoreBasedGenerativeModeling2021a}. Our model performed well at the task of class retrieval, consistently producing the desired class (Fig.~ \ref{fig:mnist}).

For our intensity-related experiment, we tested our model on its ability to generate all of the classes given different starting intensity. Because generative models struggle to extrapolate beyond training data, our model demonstrated poor performance for certain digits on initial intensity values that were too high or too low. In response to this, we picked the '1' with the highest intensity for our high-intensity test, and the '0' with the lowest for our low intensity test, as 1 had the lowest intensity of any of the numbers, and 0 had the highest. Our model performed very well on this task, consistently producing the target class even with varying intensity. See Fig.~ \ref{fig:image_applications} (d) for results. 

\begin{figure*}[ht!]
    \centering
    \includegraphics[width=1\linewidth]{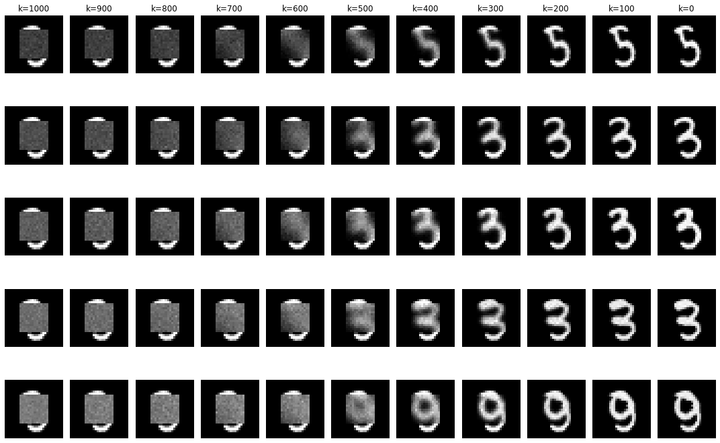}
    \caption{Unconditional MNIST inpainting with progressively increasing total intensity in the masked region. Rows correspond to increasing intensity levels, illustrating how digit identity and structure evolve under fixed spatial context.}
    \label{fig:mnist_appendix3}
\end{figure*}

In training our model to perform inpainting, we shrunk the size of the transition matrix and held the rest of the image static. We observed high quality generations very quickly, within only 40K training steps. For our intensity-related experiment, we tested the model's reaction to increasing intensity within the inpainted region and were able to see different number generations from the same starting image (Fig.~\ref{fig:mnist_appendix3} ). See section \ref{sec:comparison} for comparisons with other conditioning approaches.

\subsection{CIFAR-10 dataset}

We trained DSD on the CIFAR-10 dataset and evaluated its generative performance under different sampling configurations. Specifically, we tested four values of the Courant–Friedrichs–Lewy (CFL) tolerance parameter, $\varepsilon \in {0.01, 0.05, 0.10, 0.15}$, which controls the reverse-time integration step size. As expected, lower values of $\varepsilon$ yield more accurate reverse-time integration and lead to improved Fréchet Inception Distance (FID).

\begin{figure*}[ht!]
    \centering
    \includegraphics[width=1\linewidth]{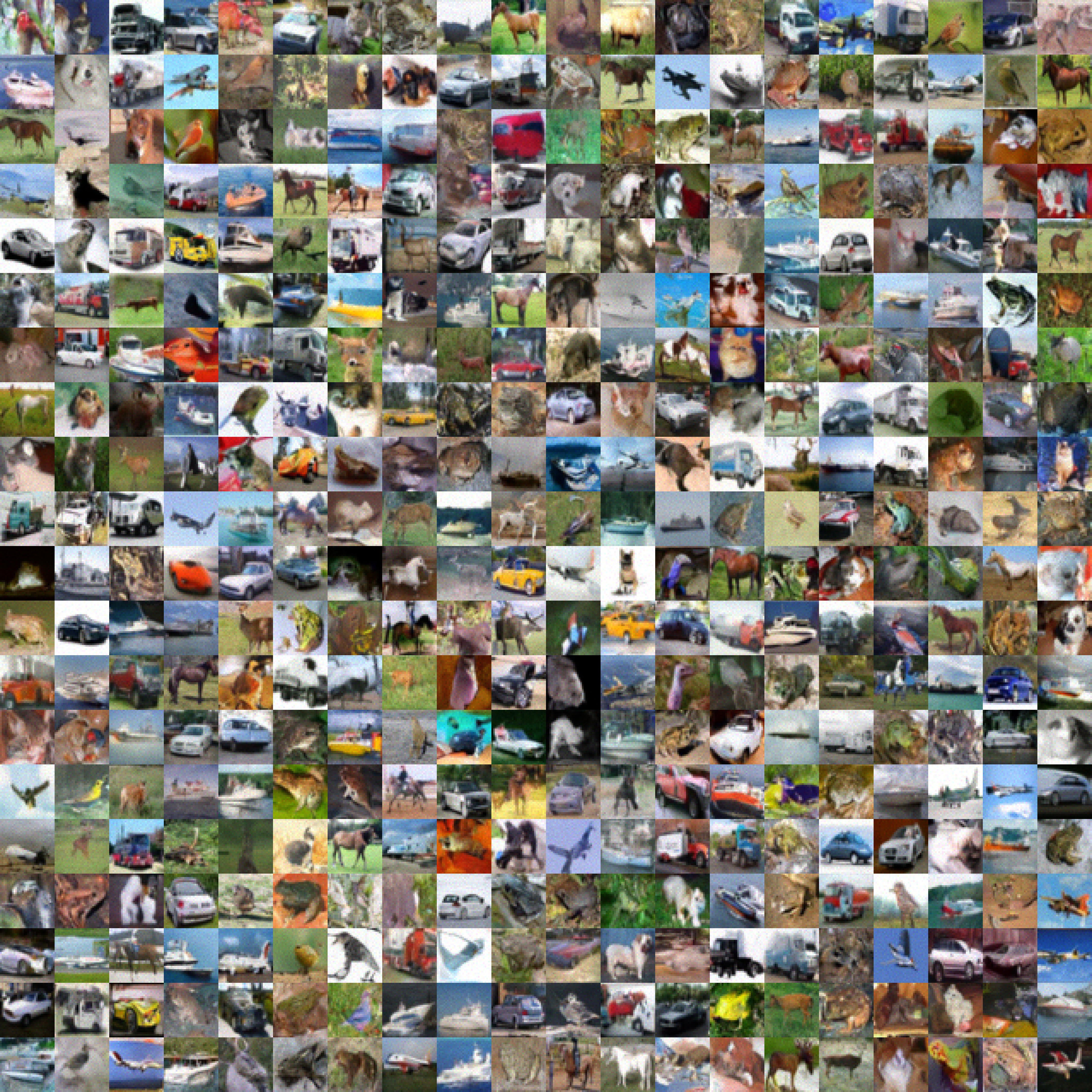}
    \caption{Unconditional CIFAR-10 generations produced by DSD with CFL tolerance $\varepsilon = 0.01$. Samples are generated without any post-processing. While structurally coherent, mild pixel-level irregularities are visible.}
    
    \label{fig:cifar-unconditional}
\end{figure*}

While FID is not the primary focus of our model, the discrete-state pixel-hoping nature of DSD introduces local high-frequency fluctuations during sampling (Fig.~\ref{fig:cifar-unconditional}) that reduce smoothness. These artifacts, though minor, negatively impact FID and perceptual metrics. To address this, we apply a lightweight bilateral filter as a post-processing step to each generated image. This filter is non-trainable, respects both spatial proximity and intensity similarity, and acts as a smooth denoising operator that preserves semantic edges and attenuates visually incoherent fluctuations. Importantly, we apply it with default hyperparameters across all images, without per-sample tuning. The filtering operation is computationally negligible and can be implemented as a fixed convolutional layer at the end of the generation pipeline.

\begin{figure*}[ht!]
    \centering
    \includegraphics[width=1\linewidth]{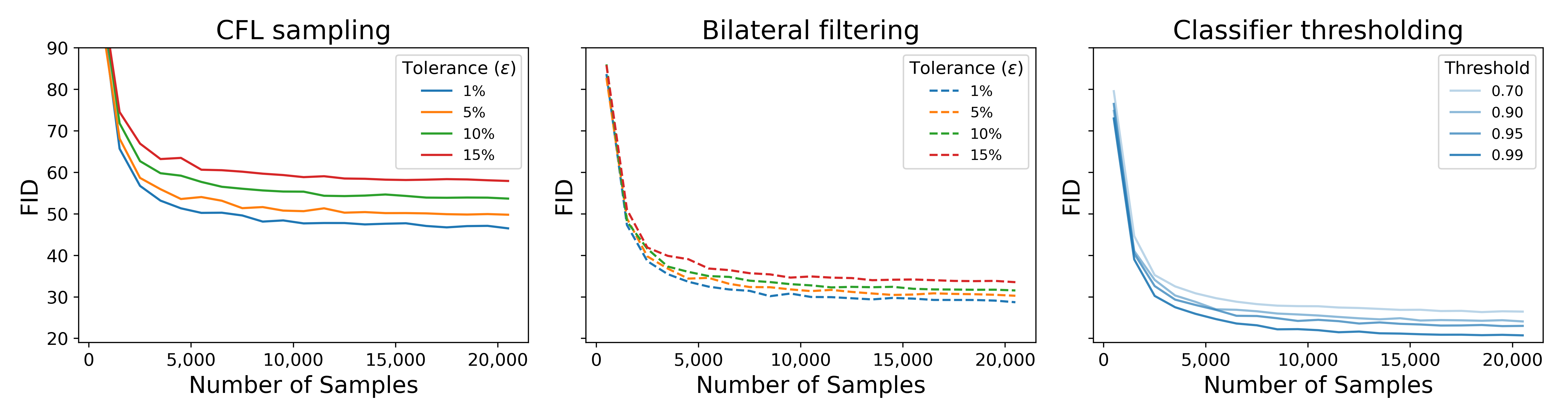}
    \caption{FID scores computed on 20,000 CIFAR-10 generations under different sampling configurations. 
\textbf{(Left:)} Samples generated with different CFL tolerances ($\varepsilon$). 
\textbf{(Middle:)} Same samples after applying a fixed bilateral filter. 
\textbf{(Right:)} Bilaterally filtered samples sampled with 1\% further refined by discarding outputs with classifier confidence below a certain threshold.}

    \label{fig:cifar-qualitative}
\end{figure*}

\begin{figure*}[ht!]
    \centering
    \includegraphics[width=1\linewidth]{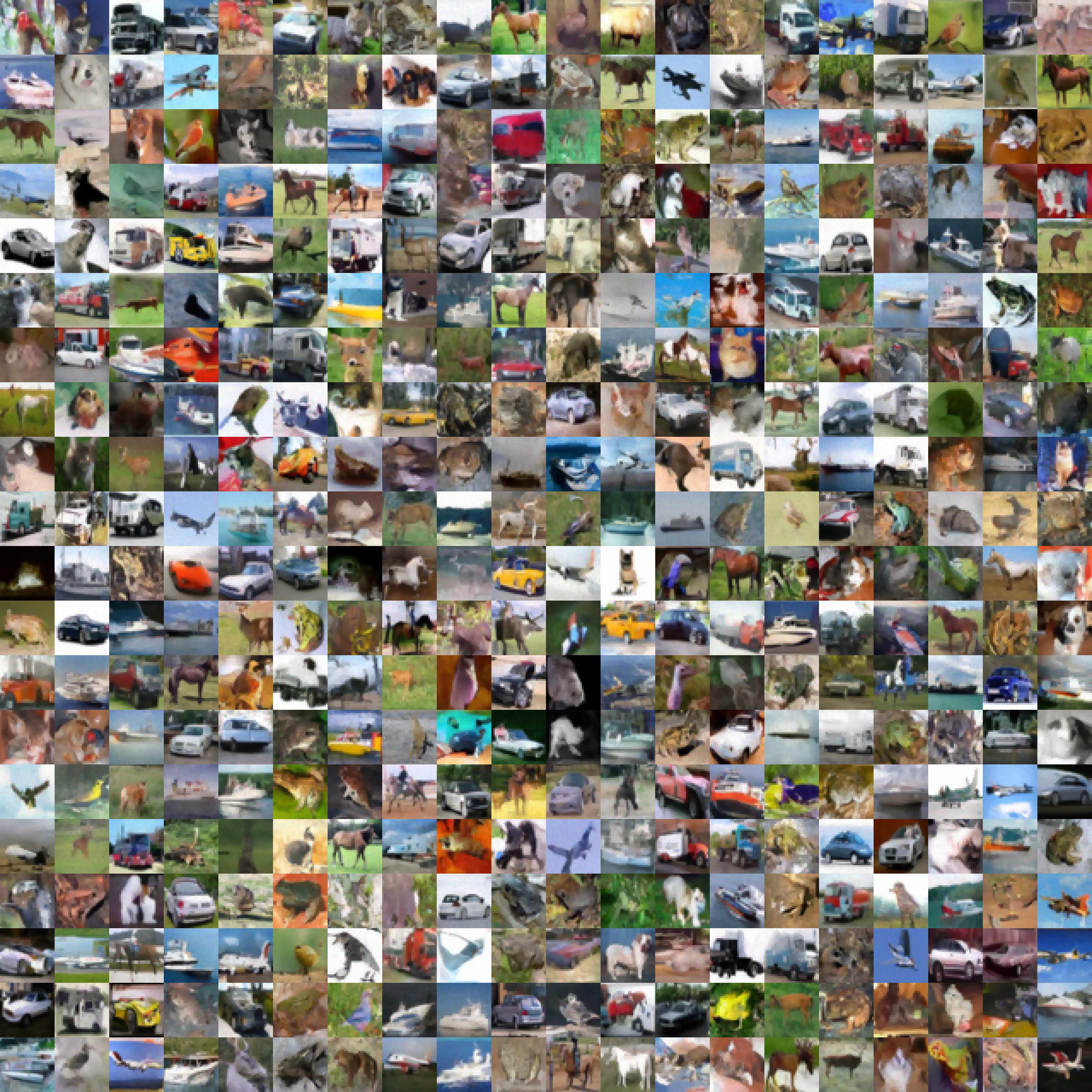}
    \caption{CIFAR-10 generations after bilateral filtering. These images correspond directly to the unfiltered samples shown in Fig.~\ref{fig:cifar-unconditional}, post-processed using a fixed, non-trainable bilateral filter. The operation smooths pixel-level irregularities introduced by the discrete sampling process while preserving structural and semantic content.}
    
    \label{fig:cifar-unconditional_filtered}
\end{figure*}

This simple step substantially improves both FID and spatial FID (sFID), without degrading the underlying structure of the samples (Fig.\ref{fig:cifar-unconditional_filtered}). In Fig.\ref{fig:cifar-qualitative}b, we quantify the impact of filtering across CFL tolerances. As an additional refinement step, we also use a pre-trained off-the-shelve CIFAR-10 classifier to discard semantically incoherent samples. Filtering based on classifier confidence can be integrated early in the sampling loop and correlates strongly with out-of-distribution or degenerate outputs (Fig.~\ref{fig:cifar-qualitative}c).

 A summary of results of these sampling experiments is shown in Table~\ref{tab:cifar10}. While an FID score of around 20 does not represent state-of-the-art performance, we hypothesize that incorporating larger architectures, advanced optimization strategies, multi-GPU trainings, and data augmentation could further reduce this gap. DSD may be a viable foundation for discrete, constraint-preserving generative modeling even in conventional RGB settings.

\begin{table}[h]
\centering
\caption{Quantitative evaluation of CIFAR-10 generations under different sampling and filtering strategies. The CFL tolerance $\varepsilon$ controls the reverse-time integration step size. Filtering refers to post-processing with a bilateral filter.}
\begin{tabular}{lccc}
\toprule
\textbf{Sampling Configuration} & \textbf{Images} &\textbf{FID} $\downarrow$ & \textbf{sFID} $\downarrow$ \\
\midrule
CFL $\varepsilon = 0.01$       &  50,000 &   46.3        &      23.4        \\ 
CFL $\varepsilon = 0.01$ + filtering  &       50,000       &       28.2      &       16.4       \\
+ Classifier thresholding ($p \geq 0.99$) & 21,780 & 20.6     &       17.7       \\
\bottomrule
\end{tabular}
\label{tab:cifar10}
\end{table}

\subsection{CelebA dataset}

\begin{figure*}[h!]
    \centering
    \includegraphics[width=.6\linewidth]{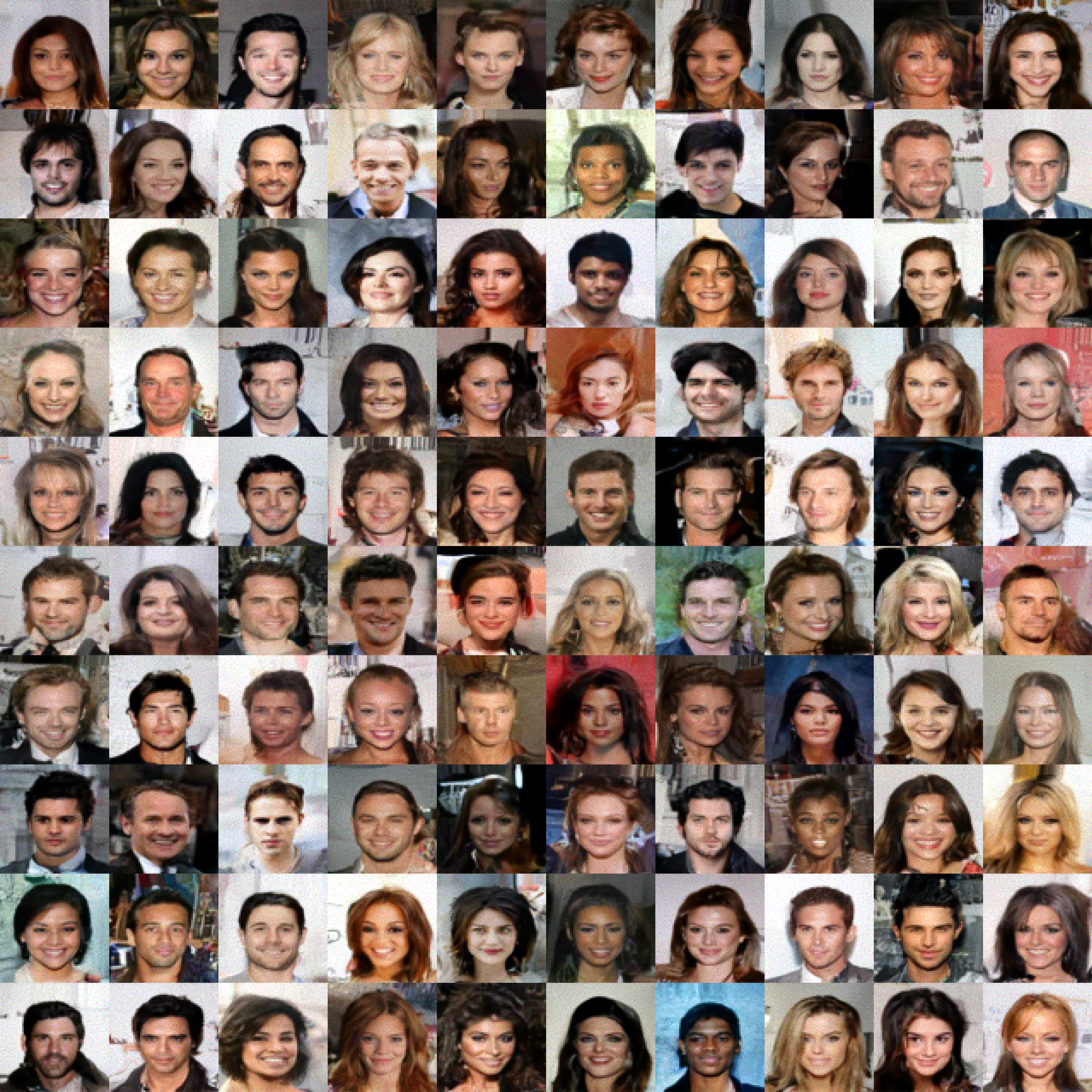}

    \caption{Unconditional generations of the CelebA 64$\times$64 dataset.}
    \label{fig:celebs}
 \end{figure*}

To demonstrate scalability to higher-resolution, human-centric data, we trained DSD on the CelebA 64$\times$64 dataset. While the resulting FID of 29 is not competitive with state-of-the-art models, it remains consistent with our CIFAR-10 results and underscores the feasibility of applying DSD to RGB images with more complex structure. As illustrated in Figure~\ref{fig:celebs}, the model is capable of producing globally coherent facial samples, despite operating in a discrete, intensity-conserving setting. We speculate that such discrete spatial models could be beneficial for downstream applications involving controlled image manipulation, such as recolorization, region editing, or consistency-preserving transformations.

\begin{table}[h]
\centering
\caption{Quantitative evaluation of CelebA 64$\times$64 generations under different sampling strategies. The CFL tolerance $\varepsilon$ controls the reverse-time integration step size. Filtering refers to post-processing with a bilateral filter. These scores were computed with 50,000 realizations.}
\begin{tabular}{lcc}
\toprule
\textbf{Sampling Configuration} & \textbf{FID} $\downarrow$ & \textbf{sFID} $\downarrow$ \\
\midrule
CFL $\varepsilon = 0.05$                 &     44.2        &      29.0        \\
CFL $\varepsilon = 0.05$ + filtering     &       28.9      &       25.7       \\
\bottomrule
\end{tabular}
\label{tab:celebA}
\end{table}

\section{Comparison with existing frameworks}
\label{sec:comparison}

\subsection{Gaussian Diffusion with Conditioning}
\label{app:comparison_Gaussian}

\begin{figure*}[ht!]
    \centering
    \includegraphics[width=1\linewidth]{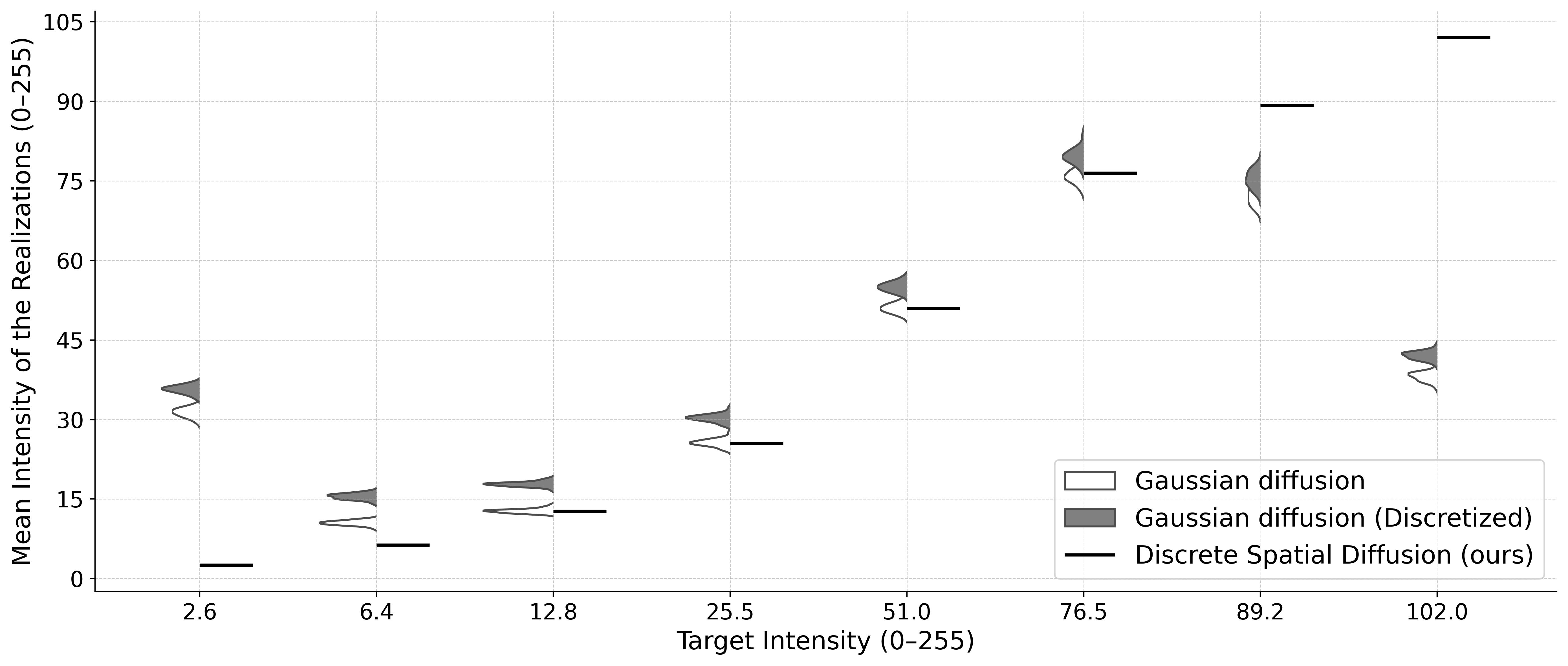}
    \caption{Sample-wise mean intensity across 1,000 realizations of mnist digits for each target intensity. We compare Gaussian diffusion \textbf{(white)}, its discretized version \textbf{(gray)}, and Discrete Spatial Diffusion \textbf{(black horizontal lines)}. Each violin shows the distribution of mean intensities computed across 1,000 individual samples.}
    \label{fig:comparison_Gaussian}
\end{figure*}

To benchmark against conventional generative models, we implemented a Gaussian diffusion baseline conditioned on the target total intensity (or mass). The architecture employed a continuous-time variant of NCSN++ with explicit conditioning on total intensity per image via an auxiliary embedding channel. Fig.~\ref{fig:comparison_Gaussian} shows the behavior of this model across a range of target intensities. Near the dataset mean, where conditioning remains within the empirical distribution, the model performs moderately well. However, closer examination reveals that the generated images often contain small negative values, even though training data was strictly bounded in $[0, 1]$. While this behavior may be inconsequential in floating-point representations, it becomes problematic when the outputs are discretized and clipped for evaluation or downstream use. In particular, thresholding negative values and rounding to 8-bit precision systematically increases the intensity error. These issues become increasingly severe in the distribution tails, where the model frequently fails to respect the specified conditioning. 

In contrast, our proposed DSD framework guarantees exact intensity preservation by construction, without requiring post hoc corrections, which holds uniformly across the intensity distribution, including extreme quantiles, as shown in Fig.~\ref{fig:comparison_Gaussian}. Notably, when sampling at the lowest intensity levels, DSD generates extremely thin or skeletal "1"s, while very high intensities produce over-saturated, bold structures such as "fat" zeros that span much of the image domain. This behavior suggests that the model is not merely enforcing conservation, but learning meaningful structural adaptations to accommodate different global constraints, consistent with underlying patterns in the data.

 Our observations indicate that: (1) intensity can be statistically conditioned within the distributional range, but without guarantee of exactness; (2) rounding and clipping of generated samples introduce further bias, revealing that the model relies on physically invalid negative intensities to approximate the constraint; and (3) extrapolation outside the data manifold results in complete failure of conditioning.

\subsection{Guidance Function Approach}

In reinforcement learning settings, a guidance function can be naturally defined using task-specific reward or cost structures \cite{janner2022planningdiffusionflexiblebehavior}. In generative modeling, however, such functions are generally unavailable, and often must be learned or heuristic in nature, as discussed in  work on posterior sampling. Specifically, exact guidance is only possible when the conditional distribution as exact posteriors (of generative samples given the constraint of interest) is known, which is not the case for total intensity constraints in image generation.

To explore whether heuristic guidance functions could serve as a surrogate, we conducted additional experiments applying a mean-squared error (MSE) penalty between the generated and target total intensity. Our results showed that while guidance can reduce error somewhat for intensities near the mean training data intensity, it fails to enforce conditioning reliably, and especially so at the tails of the  training distribution. In these cases, samples frequently severely violate the intensity constraint.

\section{Subsurface microstructures}

\subsection{Detailed description of X-ray scans of subsurface rocks}
\label{app:geology}

\begin{itemize}
    \item \textbf{Berea Sandstone}: This sandstone sample from \cite{berea}  provides a high-resolution image of the rock microstructures obtained through X-ray microtomography (X-ray $\mu$CT). In this process, the rock sample is rotated while being scanned by an X-ray beam, capturing a series of 2D radiographs at different angles. These projections are then computationally reconstructed into a 3D volume, where each voxel represents the X-ray attenuation of the material at that location.  The X-ray microtomography scans were performed using a SkyScan 1272 system, operating at 50 kV and 200 $\mu$A, with a CCD detector capturing projections at a resolution of 2.25 $\mu$m per voxel. The resulting dataset consists of grayscale images with a voxel size of 2.25 $\mu$m, where variations in intensity distinguish between the solid matrix and the pore space. The solid matrix primarily consists of tightly packed mineral grains—mostly quartz—while the pores are voids that can be occupied by fluids such as water or hydrocarbons. After preprocessing steps like contrast enhancement, noise reduction, and segmentation, the final dataset represents the  pore network. The Berea sample has a measured porosity of 18.96\% and permeability of 121 mD. This dataset is particularly useful for computational modeling, as it enables direct comparison between numerical simulations and experimentally measured permeability, providing a rich testbed for learning-based methods that seek to map complex microstructural information to macroscopic transport properties. This sedimentary rock is a well-characterized geological benchmark, widely used in fluid flow studies due to its homogeneous grain structure and consistent permeability properties, making it a good first benchmark for our study.

    \item \textbf{Savonnières Carbonate}: This carbonate sample, described in \cite{carbonate_s}, is a layered, oolithic grainstone with a wide porosity and a permeability varying from 115 to over 2000 mD, depending on local heterogeneities. The rock is characterized by a highly multimodal and interconnected pore structure, with distinct macropores and microporosity. X-ray microtomography (X-ray $\mu$CT) was used to image the sample at a resolution of 3.8~$\mu$m voxel size, revealing intricate pore geometries. The sample was scanned at the Ghent University Centre for X-ray Tomography (UGCT) using their HECTOR scanner, developed in collaboration with XRE, Belgium. The macropores include both intergranular voids and hollow ooids, while the microporosity is found within ooid shells and intergranular spaces. Micropores in the sample often serve as the primary pathways connecting poorly connected macropores, creating a complex hierarchical network. After preprocessing steps, including noise reduction, anisotropic diffusion filtering, and watershed segmentation, a multiscale pore network model was extracted. This dataset is particularly compelling due to its extreme heterogeneity, with pore sizes spanning orders of magnitude, and its ability to represent coupled serial and parallel flow pathways. Savonnières serves as a test case for studying the impact of complex samples in our workflow.
    
   \item \textbf{Massangis Limestone}: This oolitic limestone sample from \cite{carbonate_m} is a highly heterogeneous carbonate rock with a complex, multimodal pore structure resulting from diagenetic alterations, including dolomitization and dedolomitization. The rock contains a mix of intergranular and moldic macroporosity, along with microporosity concentrated in ooid rims and partially dissolved dolomite regions. Its porosity ranges from 9.5\% to 13.8\%, depending on local variations, and its permeability is highly anisotropic due to the interplay between connected macropores and poorly accessible microporosity. X-ray microtomography (X-ray $\mu$CT) was used to image the sample at a voxel resolution of 4.54 $\mu$m, capturing the intricate connectivity of macro- and micropores. The sample was scanned at the Ghent University Centre for X-ray Tomography (UGCT) using a FeinFocus FXE160.51 transmission tube, in collaboration with Paul Scherrer Institute (PSI), Switzerland. Differential imaging was applied to enhance the detection of fluid-filled microporosity, revealing  the rock’s internal heterogeneities. Unlike more uniform carbonate samples, Massangis exhibits significant spatial variations in pore connectivity, leading to zones of high permeability interspersed with isolated pore networks. This dataset serves as another challenging benchmark for modeling porous media microstructure.

\end{itemize}

\begin{figure*}[h!]
    \centering
    \includegraphics[width=1.0\linewidth]{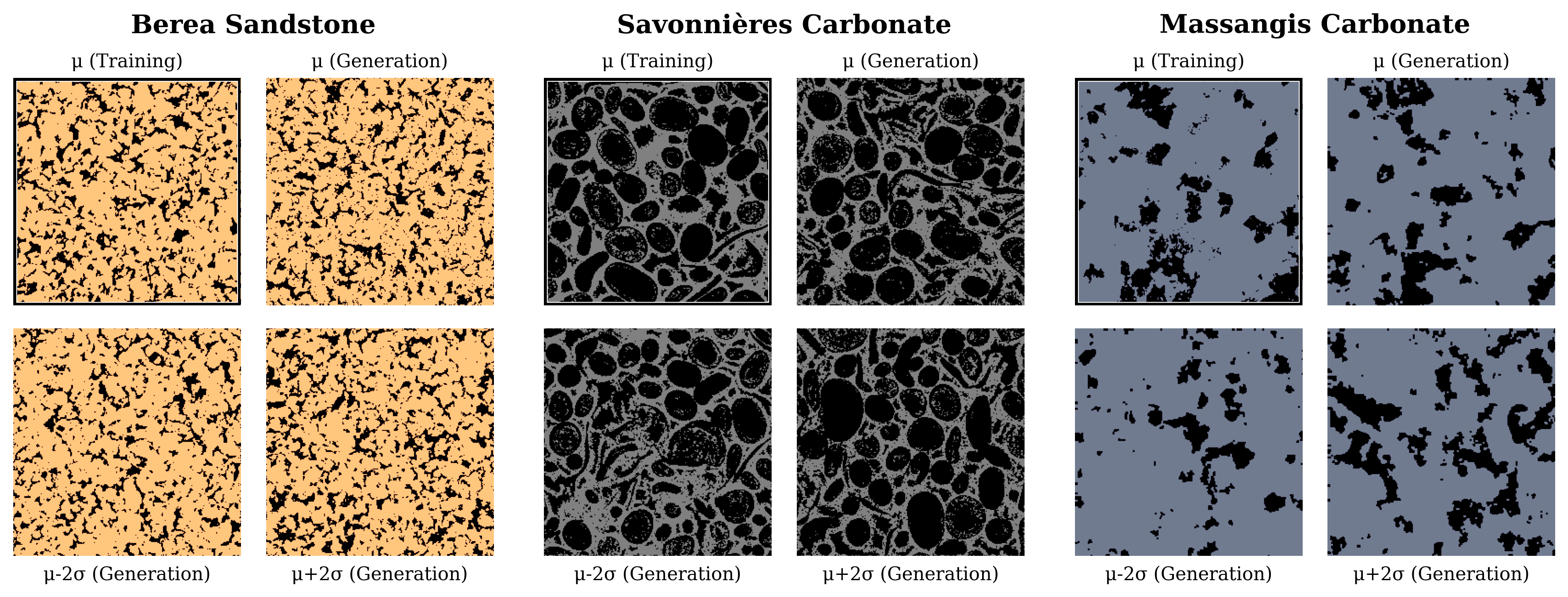}
    \vspace{-8mm}
    \caption{Schematic representation of three rock types: Berea Sandstone, Savonnières Carbonate, and Massangis Carbonate. The first image \textbf{(top left)} for each rock type shows one training sample, while the second one \textbf{(top right)} displays the generated sample conditioned on the mean intensity $\mu$ of the training set. The third \textbf{(bottom left)} and fourth \textbf{(bottom right)} samples illustrate the generated samples conditioned on $\mu-2\sigma$ and $\mu+2\sigma$, respectively, where $\sigma$ represents the standard deviation of the training set intensity distribution.}
    \label{fig:rocks_old}
 \end{figure*}

\subsection{Quantitative metrics of the generated images}
\label{app:geology_metrics}

\begin{figure}[ht!]
    \centering
    \includegraphics[width=\linewidth]{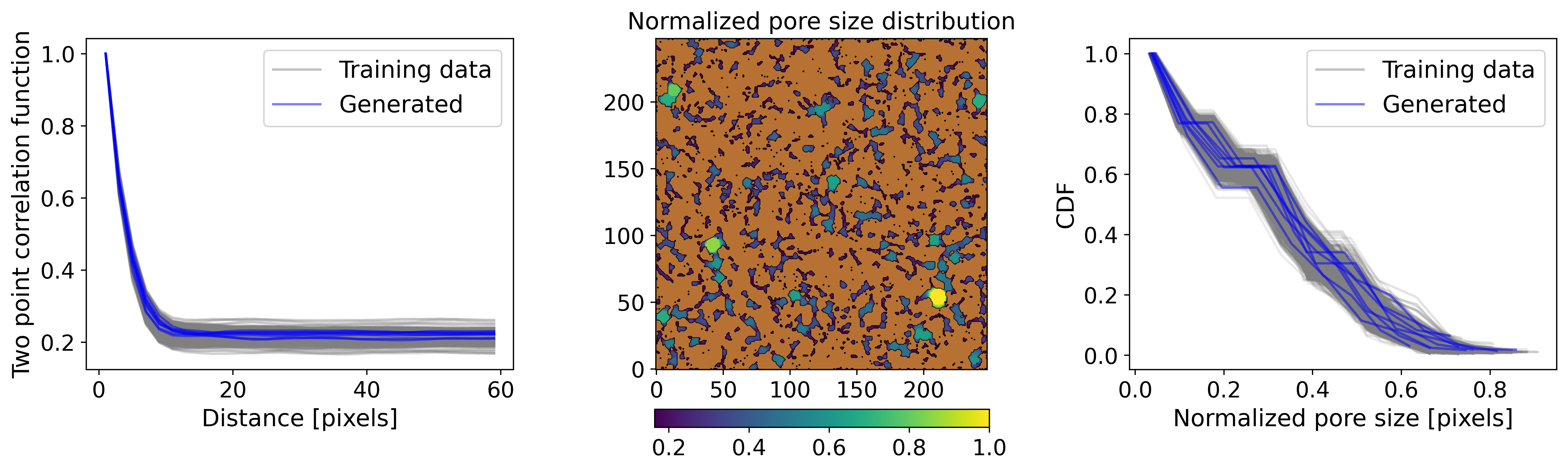}
    \caption{Quantitative comparison between training and generated rock samples. \textbf{(Left)} Two-point correlation function, showing excellent agreement of spatial features between training data (gray) and 100 randomly generated samples (blue). \textbf{(Middle)} Normalized pore size distribution schematic, with colors indicating relative pore sizes. \textbf{(Right)} Cumulative distribution function (CDF) of the normalized pore sizes, comparing the statistical distribution of training and generated samples.}
    \label{fig:pore_metrics}
\end{figure}

In porous media analysis, characterizing the spatial arrangement and size distribution of pores is crucial for understanding transport properties, mechanical behavior, and overall structure-function relationships. To quantify these characteristics, we compute the spatial correlation function and pore size distribution (PSD) using PoreSpy \cite{porespy}, a Python-based toolkit for quantitative analysis of porous media images. The Pore Size Distribution (PSD) characterizes the variation of pore sizes within a porous material, providing insights into connectivity, permeability, and flow dynamics. The most common method to determine PSD computationally is the local thickness approach. Given a binary image $I(x,y)$, where pore space is represented as 1 and solid space as 0, the pore size function $f(r)$ is defined as the probability density function (PDF) of the largest sphere that can be inscribed at any point within the pore space. The PSD provides a statistical summary of pore connectivity and transport properties. Small pores dominate permeability, while large pores govern bulk flow. Both of these metrics for the training and generated samples are shown in Fig.~\ref{fig:pore_metrics}, our realizations exhibit \textbf{excellent} agreement with the training data.

Additionally, following the evaluation protocol in \citet{lee2024microstructure}, we computed Fréchet Inception Distance (FID) for our generated carbonate samples, obtaining a score of 0.9—substantially lower than their reported value of 18.1 on rocks of similar lithology. While this comparison is not strictly one-to-one due to dataset and preprocessing differences, it offers a compelling quantitative indication of the improved realism and fidelity of DSD-generated microstructures. We note, however, that FID may have limited interpretability for binary porous media images, and should be interpreted alongside domain-specific metrics.

\subsection{Large scale generations.}
\label{sec:large_rock}

To evaluate the scalability of DSD in realistic scientific settings, we trained a model on high-resolution binary microstructure data of Leopard Sandstone, an interesting sample from the geological standpoint. This dataset consists of $1000 \times 1000$ samples representing complex subsurface structures characterized by heterogeneity, anisotropy, and nontrivial pore connectivity. Generating such large samples is particularly relevant for studying representative elementary volumes.

\begin{figure*}[h!]
    \centering
    \includegraphics[width=1.0\linewidth]{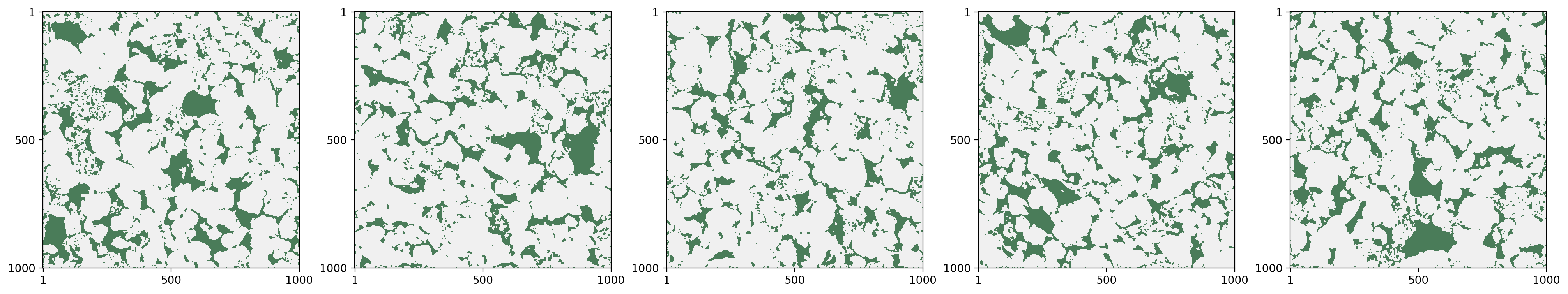}
  
    \caption{Realizations of Discrete Spatial Diffusion (DSD) on 1,000$\times$1,000 images of Leopard Sandstone. These samples demonstrate that our model scales to high-resolution, structurally complex porous media. Leopard Sandstone exhibits heterogeneous, anisotropic pore structures representative of real-world geological variability. }
    \label{fig:big_rocks}
 \end{figure*}

Despite the increased resolution and number of particles, DSD remains computationally feasible in this regime. The forward process remains efficiently parallelizable across CPU workers, and reverse-time sampling scales linearly with intensity while maintaining tractable runtimes. A subset of generated samples is shown in Fig.~\ref{fig:big_rocks}, demonstrating the model's ability to reproduce structurally diverse and geologically plausible microstructures at scale.

While extending DSD to 3D requires modest modifications to the implementation (e.g., defining a 6-connected lattice for particle jumps), we find the approach remains tractable for volumetric data, further supporting its suitability for scientific applications in porous media and materials modeling.

\section{X-ray scans of NMC cathodes}
\label{app:cathodes}
\subsection{Dataset description}
This dataset provides high-resolution 3D images of a Li-ion battery cathode composed of active material (nickel-manganese-cobalt oxide, NMC), carbon black, and a polymer binder \cite{usseglio2018resolving}. The cathode sample was imaged via X-ray microtomography (X-ray µCT) and nano-tomography (X-ray nano-CT) to capture both the overall electrode architecture and fine-scale features of the carbon/binder domain (CBD). For micro-CT, a Zeiss Xradia Versa 520 system was operated at 80 kV and 88 µA, acquiring projections at an effective isotropic voxel size of approximately 398 nm over a field of view of about 400 µm. The nano-CT scans were performed using a Zeiss Xradia Ultra 810 system with a chromium target (35 kV, 25 mA), yielding isotropic voxel sizes on the order of 126 nm across a field of view of approximately 64 µm. In both cases, the 2D radiographs were reconstructed into 3D grayscale volumes using a filtered back-projection algorithm, capturing the X-ray attenuation due to the dense NMC particles and the less attenuating pore/CBD regions.

These tomographic datasets reveal the hierarchical microstructure of the electrode, from tens-of-micrometers NMC active particles to nanometer-scale pores within the percolated carbon network. After preprocessing—such as non-local mean filtering, contrast enhancement, and slice-by-slice local thresholding—segmentation identifies three main phases: (1) the NMC active material, (2) the CBD, and (3) the pore space. Measured porosity values for these cathodes can exceed 30\%, while the typical volume fraction of active material is on the order of 40\%. The overall areal loading of the active material is around 29.78 mg·cm\textsuperscript{-2}, corresponding to about 33 mAh·cm\textsuperscript{-2} in specific capacity. These 3D reconstructions enable computational modeling of transport properties (e.g., tortuosity factor) and electrochemical performance, facilitating direct comparisons with experimentally measured parameters. Because of the electrode’s well-defined spherical NMC particles and percolating carbon network, this dataset serves as a robust benchmark for multi-scale modeling and data-driven methods that aim to link microstructural features to macroscopic cell behavior.

\subsection{Effective metrics}
The analysis of NMC cathode tomography and the generated images was conducted using three  metrics: interface length, triple-phase boundary, and relative diffusivity. These metrics are essential for quantifying the morphological and transport characteristics that influence the electrode's electrochemical performance. Below we describe these metrics in detail.

\textbf{Interface length}
refers to the total length of boundaries where two distinct phases, such as active material and pore or electrolyte, intersect. A higher interface length indicates more active sites for electrochemical reactions and enhances ion transport pathways, thereby improving the electrode’s overall performance. This metric is calculated by identifying and summing the perimeters of all phase boundaries in the segmented image.

\textbf{Triple-Phase Boundary}
denotes the regions where three different phases—typically solid active material, electrolyte, and a conductive phase or pore space—converge in the microstructure. TPBs are crucial for facilitating efficient electrochemical reactions, as they provide optimal sites where all necessary phases interact. The total TPB length is determined by locating points or lines where three phases meet and summing their lengths within the image.

\textbf{Relative Diffusivity}
quantifies the reduction in ion transport within the porous cathode structure relative to an unobstructed medium. It is defined as the ratio of the effective diffusivity, $D_{\text{eff}}$, through the porous medium to the intrinsic diffusivity, $D_0$, of the conductive phase: $D_{\text{rel}}=D_{\text{eff}}/D_0$. This reduction is primarily attributed to the geometric complexities of the microstructure, encapsulated by the tortuosity factor, $\tau$, in fact $D_{\text{rel}}=D_{\text{eff}}/D_0=V_f/\tau$, where $V_f$ is the volume fraction of the phase under analysis. 

\begin{figure*}
    \centering
    \includegraphics[width=1.0\linewidth]{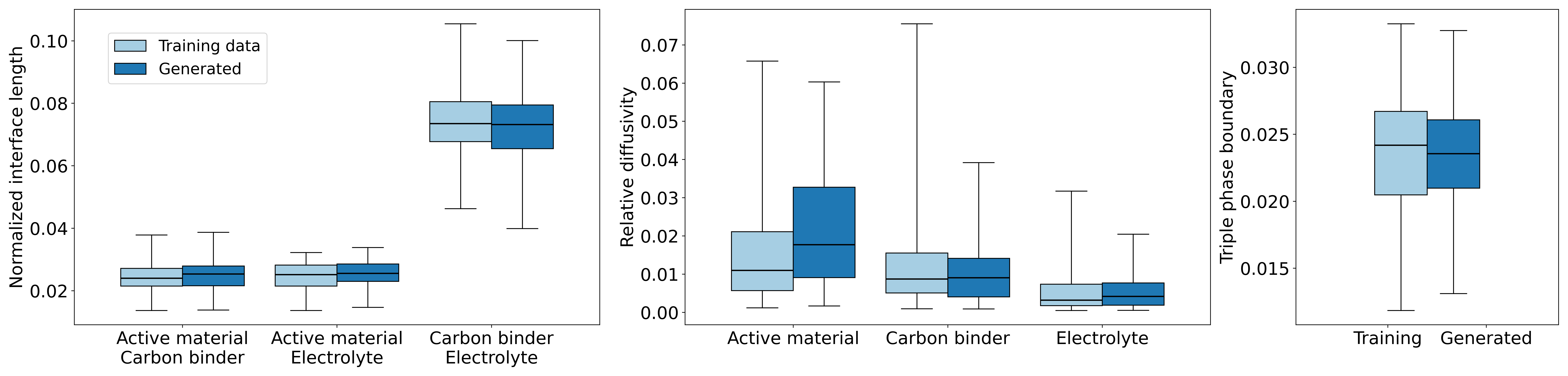}
    \caption{Microstructural characterization metrics for 80 training samples and 80 generated samples. The boxes show the 25th-50th-75th percentile, the whiskers the minimum and maximum values. Metrics computed using TauFactor~\cite{kench2023taufactor}.}
    \label{fig:electrode_samples_appendix1}
\end{figure*}

\begin{figure*}
    \centering
    \includegraphics[width=0.6\linewidth]{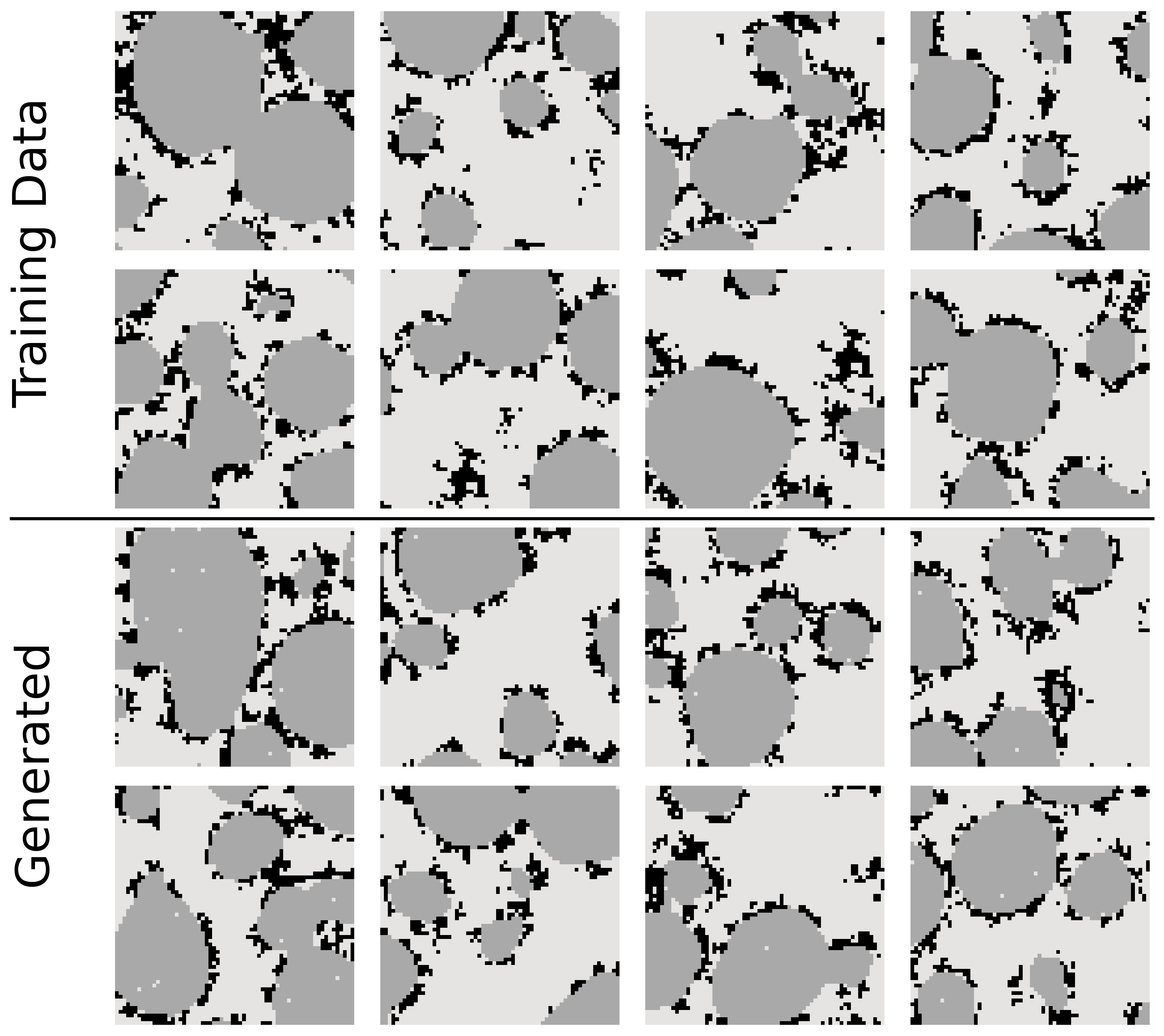}
    \caption{\textbf{(Top)} Eight randomly picked samples from the NMC cathodes dataset. \textbf{(Bottom)} Random unconditional realizations of our model.}
    \label{fig:electrode_samples_unconditional}
\end{figure*}

We computed these metrics using the Python library TauFactor \cite{kench2023taufactor}, and the comparisons between the real and generated images based on these metrics are illustrated in Fig.~\ref{fig:electrode_samples_unconditional}, while a collection of the training data and generated images is in~Fig.\ref{fig:electrode_samples_appendix1}.

\section{Hyperparameters for experiments}
\label{app:hparams}

In our experiments, we thoroughly tested our model on various hyperparameters using the MNIST dataset. The MNIST dataset was chosen as a baseline for hyperparameter testing due to its low computational training cost.  We found that our model was very robust with respect to the hyperparameters used, consistently generating quality generations without hyperparameter tuning. Due to limited compute, only limited tests were performed on CelebA, but we hypothesize that our model would perform well with different hyperparameters than the ones used. For the choice of our 'r', we chose a rate that was large enough to allow full degradation, enabling the model to learn to predict starting from full noise. See \ref{tab:hyperparams} for our hyperparameters used.

\begin{table}[!htpb]
    \centering
    \caption{Hyperparameters used in all our experiments. All models were ran in a single NVIDIA A100 (or similar).}
    \label{tab:hyperparams}
     \resizebox{0.95\linewidth}{!}{
    \begin{tabular}{lccccccc p{3cm}}
        \toprule
        &  &  & \textbf{Boundary} &  & \textbf{CFL} & \textbf{Channel} & \textbf{Training} & \\

        \textbf{Dataset} & \textbf{$r$} & \textbf{Schedule} & \textbf{Condition} & \textbf{Loss} & \textbf{Tolerance} & \textbf{Multiplier} & \textbf{Iterations} & \textbf{Notes} \\
        
        \midrule

       MNIST  & 120  & Ours  & Periodic  & Eq. \ref{eq:l1loss}  & 0.15  & (2,2,2) & 100K  & unconditional  \\
        MNIST  & 120  & Ours  & Periodic  & Eq. ~\ref{eq:likelihood-cts}  & 0.15  & (2,2,2) & 90K  & unconditional \\
        MNIST   & 120  & Ours  & No-flux  & Eq. ~\ref{eq:likelihood-cts}  & 0.15  & (2,2,2) & 80K  & unconditional \\
        MNIST & 120  & Ours  & No-flux  & Eq. ~\ref{eq:likelihood-cts}  & 0.11  & (2,2,2) & 70K  &  class-conditioned \\
        MNIST   & 85  & Ours & No-flux  & Eq. ~\ref{eq:likelihood-cts}  & 0.07  & (2,2,2) & 40K  & inpainting (14x14) \\
        CIFAR10  & 160  & Ours  & Periodic  & Eq. \ref{eq:likelihood-cts}  & Fig.~\ref{fig:cifar-qualitative}  & (1,2,2,2) & 10M  & unconditional \\
        CelebA & 200  & Ours  & No-flux  & Eq. \ref{eq:l1loss}  & 0.05  & (1,2,2,2) & 3M  &  unconditional \\
        Electrodes & 200  & $x^{5}$  & Periodic  & Eq.~\ref{eq:likelihood-cts}  & 0.01  & (1,2,2,2) & 180k  &  \\
        Rocks  & 250  & $x^{4}$  & Periodic  &  Eq.~\ref{eq:likelihood-cts}  & 0.1/0.2/0.05  & (1,2,2,2) & 50k  & tolerance avoids overlapping particles \\
        
        \bottomrule
    \end{tabular}
    }
\end{table}

\end{document}